\documentclass[12pt]{article}

\usepackage[T1]{fontenc}
\usepackage[utf8]{inputenc}
\usepackage{booktabs}
\usepackage{amsmath}
\usepackage{graphicx}
\usepackage{esint}
\usepackage[unicode=true,
 bookmarks=false,
 breaklinks=false,pdfborder={0 0 1},backref=section,colorlinks=false]
 {hyperref}

\makeatletter

\providecommand{\tabularnewline}{\\}

\newcommand{\lyxaddress}[1]{
	\par {\raggedright #1
	\vspace{1.4em}
	\noindent\par}
}

\usepackage{float}
\usepackage{units}

\DeclareMathOperator{\sgn}{sgn}

\@ifundefined{showcaptionsetup}{}{%
 \PassOptionsToPackage{caption=false}{subfig}}
\usepackage{subfig}
\makeatother

\begin{document}
\title{Density of instantaneous frequencies in the Kuramoto-Sakaguchi model}
\author{Julio D. da Fonseca$^{1}$\thanks{jcddafonseca@gmail.com}, Edson
D. Leonel$^{1}$\thanks{edson-denis.leonel@unesp.br}, and Rene O.
Medrano-T$^{1}$ $^{2}$\thanks{rene.medrano@unifesp.br} }
\maketitle

\lyxaddress{$^{1}$ Departamento de Física, Universidade Estadual Paulista, Bela
Vista, 13506-900 Rio Claro, SP, Brazil}

\lyxaddress{$^{2}$ Departamento de Física, Universidade Federal de São Paulo,
UNIFESP, 09913-030, Campus Diadema, São Paulo, Brasil}
\begin{abstract}
We obtain a formula for the statistical distribution of instantaneous
frequencies in the Kuramoto-Sakaguchi model. This work is based on
the Kuramoto-Sakaguchi's theory of globally coupled phase oscillators,
which we review in full detail by discussing its assumptions and showing
all steps behind the derivation of its main results. Our formula is
a stationary probability density function with a complex mathematical
structure, is consistent with numerical simulations and gives a description
of the stationary collective states of the Kuramoto-Sakaguchi model. 
\end{abstract}

\section{Introduction}

Synchronization is the process by which interacting oscillatory systems
adjust their frequencies in order to display the same common value
\cite{Pikovsky}. Power grids \cite{Motter13}, semiconductor laser
arrays \cite{Kozyreff}, cardiac pacemaker cells \cite{Winfree80},
and neurosciences \cite{Breakspear10} are just a few examples in
a multitude of domains where synchronization is an active research
subject.

The works of A. Winfree and Y. Kuramoto brought seminal contributions
to the study of synchronization \cite{Winfree80,Winfree67,KuramotoA,KuramotoB}.
Inspired by Winfree's pioneering ideas \cite{Kuramoto-video}, Kuramoto
formulated a model of coupled phase oscillators today known as \emph{Kuramoto
model}. The Kuramoto model was introduced in Ref. \cite{KuramotoA},
and its first and more detailed analysis by Kuramoto himself, published
in Ref. \cite{KuramotoB}. Since then, many studies about the Kuramoto
model and its variants appeared in the literature. (Reviews about
the Kuramoto model can be found in Refs. \cite{Strogatz00,daFonseca18};
see Refs. \cite{Acebron,Gupta,Rodrigues,Mihara2022a} for later studies
related to variants of the Kuramoto model and their applications.)

The Kuramoto model consists of an ensemble of oscillators with a mean-field
coupling and randomly distributed \emph{natural (or intrinsic) frequencies}.
An oscillator is characterized by its \emph{phase}, and the first-order
time derivative of the oscillator's phase, which here we call \emph{instantaneous
frequency}, is defined by an autonomous first-order ordinary differential
equation. The theoretical analysis of the Kuramoto model \cite{KuramotoB,daFonseca18}
evinces a transition between two stationary collective states: an
\emph{incoherent} state and a synchronization one. In the incoherent
state, instantaneous and natural frequencies have the same statistical
distribution. In the synchronization state, some oscillators have
instantaneous frequencies sharing the same value. The number of synchronized
oscillators depends on the model's parameter called \emph{coupling-strength,
}and synchronization only occurs for a coupling strength above a critical
value \cite{KuramotoB,daFonseca18}. In a simplified version of the
Kuramoto model, identical (with the same natural frequencies) and
symmetrically-coupled oscillators show multiple regular attractors
\cite{Mihara2019}, and the synchronization state is the most probable
one\cite{Girvan2016,Mihara2022b}.

H. Sakaguchi and Y. Kuramoto created a generalization of the Kuramoto
model \cite{Sakaguchi} introducing into the coupling function a phase
shift, also called \emph{phase-lag. }The Kuramoto-Sakaguchi model
and its variants appear in the study of a wide range subjects such
as chimera states \cite{Abrams,Laing}, chaotic transients \cite{Wolfrum},
pulse-coupled oscillators \cite{Pazo}, and Josephson-junction arrays
\cite{Wiesenfeld}. In addition, the coupling function with a phase-lag
can be interpreted as an approximate model of interactions with time-delayed
phases \cite{Crook}. The Kuramoto-Sakaguchi model exhibits the same
stationary collective states as the original Kuramoto model \cite{Sakaguchi}.

Collective states of Kuramoto-like models are commonly characterized
by means of an \emph{order parameter}, which is zero in the incoherent
state and takes finite values in the synchronization state. In this
work, we follow a different approach from the usual order-parameter
analysis uncovering how instantaneous frequencies are statistically
distributed in the stationary collective states of the Kuramoto-Sakaguchi
model. Instantaneous frequencies collectively reflect the occurence
of synchronized behavior \cite{Fonseca20}, and they are also relevant
in the study of other phenomena (e.g. frequency spirals \cite{Strogatz16}).

We will show how to obtain a formula for the statistical distribution
of instantaneous frequencies. The formula is defined by a stationary
probability density function, which we refer to as \emph{density of
instantaneous frequencies}.\emph{ }Our goal is similar to the one
pursued in Ref. \cite{Fonseca20} for the Kuramoto model, but here
we will show how to obtain a more general result in a more straightforward
way. A related (but still rather a different) problem was addressed
by Sakaguchi and Kuramoto in Ref. \cite{Sakaguchi}, where they analyzed
the statistical distribution of \emph{coupling-modified frequencies},
namely instantaneous frequencies averaged over infinitely large time
intervals (see Refs. \cite{daFonseca18,Fonseca20} for further details).

This work is based on the Kuramoto-Sakaguchi theory, described, as
far as we know, \emph{only} in Ref. \cite{Sakaguchi}. We will discuss
the fundamental assumptions of the Kuramoto-Sakaguchi theory and detail
how its main results can be derived. Our opinion is that an explicit
presentation of the Kuramoto-Sakaguchi theory is still absent.

We organized this paper as follows. In Section \ref{sec:KS-theory},
we present the Kuramoto-Sakaguchi theory and \emph{state diagrams}
pointing out the transition between the incoherent and synchronization
states. In Section \ref{sec:G}, we extend the Kuramoto-Sakaguchi
theory by providing additional analytical results and obtaining the
formula of the density of instantaneous frequencies. In Section \ref{sec:Application},
we discuss the properties of our formula in a specific application
example, in which natural frequencies have a Gaussian statistical
distribution. In Section \ref{sec:num}, we check the consistency
of our formula with numerical simulation data. Conclusions and an
outlook on possible research directions are given in Section \ref{sec:Conclusion}.

\section{Kuramoto-Sakaguchi theory \label{sec:KS-theory}}

The Kuramoto-Sakaguchi (KS) model\cite{Sakaguchi} consists of an
infinitely large number $N$ of all-to-all coupled oscillators. The
state of an oscillator of index $i=1...N$ is characterized by its
phase $\theta_{i}$, which changes in time according to 
\begin{equation}
\dot{\theta_{i}}=\omega_{i}-\frac{K}{N}\sum_{j=1}^{N}\sin(\theta_{i}-\theta_{j}+\alpha),\label{eq:KS-model}
\end{equation}
where $\dot{\theta_{i}}$ is the first-order time-derivative of $\theta_{i}$,
$\omega_{i}$ is a random number with a prescribed density\footnote{For the sake of simplicity, hereafter we always use the term ``density''
to refer to a probability density function. }, and $K$ and $\alpha$ are real constant parameters. \emph{We refer
to $\dot{\theta_{i}}$ as the instantaneous frequency, $\omega_{i}$
is called natural frequency and $K$, the coupling strength. } The
oscillator of index $j$ can be represented by the complex number
$\exp(i\theta_{j})$. Oscillators are then points in a complex plane
moving over a unit-radius circle centered at the origin.

A valuable concept for the analysis of collective behavior in the
KS model is that of \emph{mean field}, defined by

\begin{equation}
Z=\frac{1}{N}\sum_{j=1}^{N}\exp(i\theta_{j}),\label{eq:mean_oscillator}
\end{equation}
which can be interpreted as the \emph{average oscillator state}. The
mean field can be written as a complex number 
\begin{equation}
Z=R\exp(i\Theta)\label{eq:mean-cn}
\end{equation}
where $\Theta$ denotes the mean-field phase and $R$, the mean-field
modulus, referred to as \emph{order parameter}. If the oscillators
are quasi-aligned, i.e., they have approximately the same phase, then
$R\simeq1$. Yet, a quasi-uniform scattering of all oscillator-points
over the circle results in a mean-field located near the origin, i.e.
$R\simeq0$.

For $N\longrightarrow\infty$, the mean field, at a time instant $t$,
is given by

\begin{equation}
Z=\intop_{-\pi}^{+\pi}\exp(i\theta)n(\theta,t)d\theta,\label{eq:expected_oscillator}
\end{equation}
where $n(\theta,t)$ is the density of phases at the same time instant.
\emph{Two simple scenarios are assumed concerning the properties of
$n(\theta,t)$ in the long-time ($t\longrightarrow\infty$) and large-size
($N\longrightarrow\infty$) limits.} First, $n(\theta,t)=\frac{1}{2\pi}$
for $-\pi<\theta\leq+\pi$, and $n(\theta,t)=0$, otherwise, i.e.
$n(\theta,t)$ is a time-independent and uniform density (the value
$\frac{1}{2\pi}$ comes from the normalization condition). Second,
$n(\theta,t)$ is a steadily traveling wave with velocity $\Omega$,
i.e. $n(\theta-\Omega\Delta t,t)=n(\theta,t+\Delta t)$ for any time
instant $t$ and time interval $\Delta t$. This means that the wave
profile does not change in time, and the wave propagates with constant-in-time
velocity $\Omega$. We call the wave-propagation velocity, $\Omega$,
\emph{synchronization frequency}. 

The first scenario defines the \emph{incoherent state}, and the second,
the \emph{synchronization state}. In the incoherent state, oscillators
are uniformly spread over the unit circle. In the synchronization
state, a bunch of oscillators is \emph{synchronized}, that is, they
change collectively their phase at the same constant rate $\ensuremath{\Omega}.$

After inserting a uniform phase density in Eq. (\ref{eq:expected_oscillator}),
we see that $Z=0$. So, from Eq. (\ref{eq:mean-cn}), \emph{the order
parameter ($R$) is zero in the incoherent state. }Yet, if a KS system
exhibits synchronization, then \emph{the assumption of a traveling
wave with a stationary and non-uniform profile, moving with constant
velocity $\Omega$, means that $R$ is finite and time-independent.}
Moreover, $Z$ moves in the complex plane following a circular path
of radius $R$ and velocity $\Omega$. That is, a uniform circular
motion given by 
\begin{equation}
Z(t)=R\exp\left[i\left(\Omega t+\Theta_{0}\right)\right],\label{eq:Z-circle}
\end{equation}
where $\Theta_{0}$ is the mean-field phase at an arbitrary initial
time instant.

Let us consider KS oscillators in a different complex plane, with
the same origin as the previous one, but with both axis rotating with
angular velocity $\Omega$. In the new rotating frame, we represent
the oscillator of index $j$ by the complex number $\exp(i\psi_{j})$,
where $\psi_{j}$ is the oscillator's phase. The analogous of Eqs.
(\ref{eq:mean_oscillator}), (\ref{eq:mean-cn}), and (\ref{eq:expected_oscillator})
are 
\begin{equation}
Z^{'}=\frac{1}{N}\sum_{j=1}^{N}\exp(i\psi_{j}),\label{eq:mean_oscillator_new_frame}
\end{equation}

\begin{equation}
Z'=R\exp(i\Psi),\label{eq:mean-cn-new-frame}
\end{equation}
and, for $N\longrightarrow\infty$, 
\begin{equation}
Z^{'}=\intop_{-\pi}^{+\pi}\exp(i\psi)n(\psi)d\psi.\label{eq:expected_oscillator_new_frame}
\end{equation}

The quantities $Z^{'}$ and $n(\psi)$ are representations of the
mean field and the phase density in the rotating frame. Comparing
Eq. (\ref{eq:mean-cn}) to Eq. (\ref{eq:mean-cn-new-frame}), we see
that $Z$ and $Z'$ have the same length $R$. The mean-field length
is invariant to the change of frames because the phase density profile
is kept unchanged. Note also that, in Eq. (\ref{eq:expected_oscillator_new_frame}),
we removed the time dependence from the phase density, since both
the rotating frame and the steadily traveling wave move together with
the same velocity $\Omega$. So, both $R$ and $\Psi$ are time-independent,
which is the same as stating that \emph{the mean field is fixed in
the rotating frame}.

Some conventions are useful to simplify the analysis of the KS model
at a time instant $t>t_{o}=0$, with $t_{0}$ denoting the initial
time instant. We choose a fixed frame such that its real axis has
the same direction as the mean field at time $t_{0}$. So, $\Theta_{0}=0$
and $Z(0)=R$. Another important convention is defining a rotating
frame such that, at the initial time $t_{0}$, its real axis is dephased
by $\alpha$ from the fixed-frame's real axis (the same parameter
$\alpha$ of Eq. (\ref{eq:KS-model})). This is the same as setting
$\Psi=\alpha$.

Thus, from Eqs. (\ref{eq:Z-circle}) and (\ref{eq:mean-cn-new-frame}),
at the time instant $t$, 
\begin{equation}
Z(t)=R\exp\left(i\Omega t\right)\label{eq:mean-cn-fixed-frame}
\end{equation}
and 
\begin{equation}
Z'=R\exp i\alpha.\label{eq:mean-cn-new-frame-alpha}
\end{equation}
Also, as a consequence of the above conventions, a simple geometric
inquiring yields the relations 
\begin{equation}
\dot{\psi_{i}}=\dot{\theta_{i}}-\Omega\label{eq:psi-dot-def}
\end{equation}
 and 
\begin{equation}
\psi_{i}=\theta_{i}-\Omega t+\alpha.\label{eq:psi-def}
\end{equation}
In Eq. (\ref{eq:psi-dot-def}), $\dot{\psi_{i}}$ is the instantaneous
frequency of an oscillator of index $i$ in the rotating frame.

Using Eqs. (\ref{eq:psi-dot-def}) and (\ref{eq:psi-def}), we can
recast Eq. (\ref{eq:KS-model}) as 
\begin{equation}
\dot{\psi_{i}}=\omega_{i}-\Omega-\frac{K}{N}\sum_{j=1}^{N}\sin(\psi_{i}-\psi_{j}+\alpha).\label{eq:Kuramoto_model_rot}
\end{equation}
Multiplying the right-hand sides of Eqs. (\ref{eq:mean_oscillator_new_frame})
and (\ref{eq:mean-cn-new-frame-alpha}) by $\exp\left[-i\left(\psi_{i}+\alpha\right)\right]$
and equating their imaginary parts result in 
\begin{equation}
N^{-1}\sum_{j=1}^{N}\sin(\psi_{i}-\psi_{j}+\alpha)=R\sin\psi_{i}.\label{eq:aux}
\end{equation}
Substituting the summation in Eq. (\ref{eq:Kuramoto_model_rot}) by
the right-hand-side of Eq. (\ref{eq:aux}) gives 
\begin{equation}
\dot{\psi_{i}}=\omega_{i}-\Omega-KR\sin\psi_{i},\label{eq:KS-KR}
\end{equation}
which is a simple formulation of the KS model in the rotating frame.
We emphasize that $\omega_{i}$, $K$, $\Omega$, and $R$ are constant-in-time
numbers: $\omega_{i}$ is a sample from a random variable with a given
probability density function $g$; $K$ is a given coupling strength;
$\Omega$ and $R$ are constants to be determined.

The differential equation (\ref{eq:KS-KR}) gives a more detailed
picture of synchronization in the KS model. For $\left|\omega_{i}-\Omega\right|>KR$,
Eq. (\ref{eq:KS-KR}) has no equilibrium point. When $\left|\omega_{i}-\Omega\right|=KR$,
a pair of stable and unstable equilibria emerges by a fold bifurcation,
and they become apart as $\left|\omega_{i}-\Omega\right|<KR$. Since
the phase domain is closed ($\left|\psi_{i}\right|\leq\pi$), the
phase of an oscillator $i$, for which $\left|\omega_{i}-\Omega\right|\leq KR$,
will always converge to an attractor (stable equilibrium point) defined
by
\begin{equation}
\psi_{i}^{*}=\sin^{-1}\left(\frac{\omega_{i}-\Omega}{KR}\right),\label{eq:attractor}
\end{equation}
where $\sin^{-1}$ is an inverse of the $sin$ function with domain
$\left[-1,+1\right]$ and image $\left[-\frac{\pi}{2},+\frac{\pi}{2}\right]$.
Then, $-\frac{\pi}{2}\leq\psi_{i}^{*}\leq+\frac{\pi}{2}$.

The latter case, where oscillator $i$ has a natural frequency such
that $-KR+\Omega\leq\omega_{i}\leq\Omega+KR$, means, according to
Eqs. (\ref{eq:psi-dot-def}), (\ref{eq:psi-def}), (\ref{eq:KS-KR}),
and (\ref{eq:attractor}), that $\psi_{i}\longrightarrow\psi_{i}^{*}$,
$\theta_{i}\longrightarrow\psi_{i}^{*}+\Omega t-\alpha$, $\dot{\psi_{i}}\longrightarrow0$,
and $\dot{\theta_{i}}\longrightarrow\Omega$ as $t\longrightarrow+\infty$.
This is the case of a synchronized oscillator, or, following Kuramoto's
terminology, an \emph{S oscillator. }But, for $\left|\omega_{i}-\Omega\right|>KR$,
i.e, oscillator $i$ has a natural frequency such that $\omega_{i}<-KR+\Omega$
or $\Omega+KR<\omega_{i},$ Eq. (\ref{eq:KS-KR}) has no equilibrium
point. Then, oscillator-$i$'s phase varies according to Eq. (\ref{eq:KS-KR})
without slowing down towards an asymptotic value. This is a desynchronized
oscillator, or, simply, a \emph{D oscillator}.

Let $n(\psi,\omega)$ denote the \emph{joint density} for a rotating-frame
phase $\psi$ and a natural frequency $\omega$. The associated \emph{marginal
phase density} is given by 
\begin{equation}
n(\psi)=\intop_{-\infty}^{+\infty}n(\psi,\omega)\,d\omega,\label{eq:marginal}
\end{equation}
where $-\pi<\psi\leq+\pi$. Eq. (\ref{eq:marginal}) can be rewritten
as

\begin{equation}
n(\psi)=\intop_{\left|\omega-\Omega\right|\leq KR}n(\psi,\omega)\,d\omega+\intop_{\left|\omega-\Omega\right|>KR}n(\psi,\omega)\,d\omega.\label{eq:marginal-decomposed}
\end{equation}
The first term in the right-hand side of Eq. (\ref{eq:marginal-decomposed})
is the phase density for S oscillators, and the second one, the phase
density for D oscillators. The two terms are functions of $\psi$
which we denote by $n_{S}(\psi)$ and $n_{D}(\psi)$, respectively.
Then, 
\begin{equation}
n(\psi)=n_{S}(\psi)+n_{D}(\psi).\label{eq:n}
\end{equation}

Let $n(\psi|\omega)$ be the \emph{conditional phase density} for
a given natural frequency $\omega$. If $g(\omega)$ is the natural-frequency
density, replacing $n(\psi,\omega)$ with $n(\psi|\omega)g(\omega)$
leads to
\begin{equation}
n_{S}(\psi)=\intop_{\Omega-KR}^{\Omega+KR}n_{S}(\psi|\omega)g(\omega)\,d\omega\label{eq:ns}
\end{equation}
and 
\begin{equation}
n_{D}(\psi)=\intop_{-\infty}^{\Omega-KR}n_{D}(\psi|\omega)g(\omega)\,d\omega+\intop_{\Omega+KR}^{+\infty}n_{D}(\psi|\omega)g(\omega)\,d\omega,\label{eq:nd}
\end{equation}
where we use the following definitions: $n_{S}(\psi|\omega)\equiv n(\psi|\omega)$
for $\left|\omega-\Omega\right|\leq KR$, and $n_{D}(\psi|\omega)\equiv n(\psi|\omega)$
for $\left|\omega-\Omega\right|>KR$. Thus, to find expressions for
$n_{S}(\psi)$ and $n_{D}(\psi)$, we have to determine $n_{S}(\psi|\omega)$
and $n_{D}(\psi|\omega)$.

For a generic S oscillator with natural frequency $\omega$, Eq. (\ref{eq:attractor})
has the alternative form 
\begin{equation}
\psi^{*}(\omega)=\sin^{-1}\left(\frac{\omega-\Omega}{KR}\right).\label{eq:attractor-1}
\end{equation}
Since $\psi^{*}(\omega)$ is an attractor, the phase of this oscillator
is always in an arbitrarily small neighborhood of \textbf{$\psi^{*}(\omega)$
}for a sufficiently long time\textbf{.} Then, 
\begin{align}
\intop_{I}n_{S}(\psi|\omega)\,d\psi= & \begin{cases}
1,\quad & \psi^{*}(\omega)\in I\\
0, & \psi^{*}(\omega)\notin I
\end{cases}\label{eq:ps-cond-integral}
\end{align}
where $I$ is an arbitrary interval contained in $\left(-\pi,+\pi\right]$.
Eq. (\ref{eq:ps-cond-integral}) is the same as stating that 
\begin{equation}
n_{S}(\psi|\omega)=\delta\left[\psi-\psi^{*}\left(\omega\right)\right].\label{eq:ns-cond}
\end{equation}
We emphasize that Eq. (\ref{eq:ns-cond}) holds only for $\left|\omega-\Omega\right|\leq KR$
or, equivalently, $-KR+\Omega\leq\omega\leq\Omega+KR$.

Using (\ref{eq:attractor-1}) and (\ref{eq:ns-cond}) to solve the
integral in (\ref{eq:ns}), we obtain 
\begin{align}
n_{S}(\psi)= & \begin{cases}
g(\Omega+KR\sin\psi)KR\cos\psi,\quad & \left|\psi\right|\leq\frac{\pi}{2}\\
0, & \left|\psi\right|>\frac{\pi}{2}
\end{cases}\label{eq:ns-1}
\end{align}
According to Eq. (\ref{eq:ns-1}), $n_{S}(\psi)\longrightarrow0$
as $R\longrightarrow0^{+}$, that is, the number of S oscillators
goes to zero if the order parameter becomes small by varying $K$
and $\alpha$. Moreover, if $R$ is finite, then $n_{S}(\psi)=0$
for $\left|\psi\right|>\frac{\pi}{2}$ and $n_{S}(\psi)>0$ for $\left|\psi\right|\leq\frac{\pi}{2}$.
This comes from the property that, for a sufficiently long time, S-oscillator
phases are arbitrarily near their respective attractors, which belong,
all of them, to the interval $\left[-\frac{\pi}{2},+\frac{\pi}{2}\right]$
(See Eq. (\ref{eq:attractor})).

As mentioned earlier, finding a formula for $n_{D}(\psi)$ requires
finding a formula for $n_{D}(\psi|\omega).$ This can be done by considering
a small \emph{control interval }contained in the phase domain and
where the time-variation of the number of D oscillators with a given
natural frequency is balanced with the flow of the same type of oscillators
into and out from the interval. Defining the control interval by $[\psi,\psi+\delta\psi]$,
we have 
\begin{equation}
\partial_{t}\intop_{\psi}^{\psi+\delta\psi}n_{D}(\psi'|\omega)d\psi'=n_{D}(\psi|\omega)\dot{\psi}(\psi)-n_{D}(\psi+\delta\psi|\omega)\dot{\psi}(\psi+\delta\psi),\label{eq:conservation}
\end{equation}
where $\dot{\psi}(\psi)$ and $\dot{\psi}(\psi+\delta\psi)$ are the
rotating-frame instantaneous frequencies at phases $\psi$ and $\psi+\delta\psi$
for D oscillators with a given natural frequency $\omega$. The rotating-frame
instantaneous frequency for a D oscillator can be defined by 
\begin{equation}
\dot{\psi}(\psi)=\omega-\Omega-KR\sin\psi,\label{eq:D-f}
\end{equation}
which is the same as (\ref{eq:KS-KR}) without the index notation
but with the condition that $\left|\omega-\Omega\right|>KR$. In Eq.
(\ref{eq:conservation}), on the left-hand side is the time-variation
of the probability of finding D oscillators in the interval $[\psi,\psi+\delta\psi]$
with a given natural frequency $\omega$. The probability flow at
the endpoints of the same interval, namely $\psi$ and $\psi+\delta\psi$,
is given by the right-hand side of (\ref{eq:conservation}).

By expanding both $n_{D}(\psi'|\omega)$ and $n_{D}(\psi+\delta\psi|\omega)\dot{\psi}(\psi+\delta\psi)$
near $\psi$, taking the limit $\delta\psi\longrightarrow0^{+}$,
and neglecting high-order terms, we obtain the continuity equation
\begin{equation}
\partial_{t}n_{D}(\psi|\omega)+\partial_{\psi}[n_{D}(\psi|\omega)\dot{\psi}]=0.\label{eq:continuity}
\end{equation}
We remind the reader that $\left|\omega-\Omega\right|>KR$ for the
given value of $\omega$ in Eq (\ref{eq:continuity}).

\emph{An important assumption in KS theory is to consider that $n_{D}(\psi|\omega)$
is a stationary density.} This is consistent with the previously discussed
assumption of stationarity for $n(\psi)$. Since $n(\psi)=n_{S}(\psi)+n_{D}(\psi)$
and $n_{S}(\psi)$ are both stationary, $n_{D}(\psi)$ should also
be. According to Eq. (\ref{eq:nd}), the simplest way to accomplish
a stationary $n_{D}(\psi)$ is by assuming that $n_{D}(\psi|\omega)$
is also stationary. So, from $\partial_{t}n_{D}(\psi|\omega)=0$ and
(\ref{eq:continuity}), 
\begin{equation}
n_{D}(\psi|\omega)=\frac{C(\omega)}{\dot{\psi}(\psi)},\label{eq:nd-cond-with-C}
\end{equation}
where $C(\omega)$ is a constant with respect to $\psi$ but possibly
depending on $\omega$. Eq. (\ref{eq:nd-cond-with-C}) means that
$D$ oscillators accumulate at phases with low variation rates (low
instantaneous frequencies) and are less probable to be found at phases
with high variation rates (high instantaneous frequencies).

Applying the normalization condition to both sides of (\ref{eq:nd-cond-with-C})
results in 
\begin{equation}
\frac{1}{C(\omega)}=\intop_{-\pi}^{\pi}\frac{d\psi'}{\dot{\psi}(\psi')}.\label{eq:C}
\end{equation}
Using Eq. (\ref{eq:D-f}) and formula (2.551-3) from Ref. \cite{Gradshteyn},
we can solve the integral in Eq. (\ref{eq:C}) and obtain 
\begin{equation}
C(\omega)=\begin{cases}
C_{+}(\omega),\quad & \omega>\Omega+KR\\
C_{-}(\omega), & \omega<\Omega-KR
\end{cases}\label{eq:C+-}
\end{equation}
where 
\begin{equation}
C_{\pm}(\omega)=\pm\frac{\sqrt{(\omega-\Omega)^{2}-(KR)^{2}}}{2\pi}.\label{eq:C+-2}
\end{equation}

By inspecting Eqs. (\ref{eq:C}) and (\ref{eq:C+-}), we get a simple
interpretation of the quantity $C(\omega)$. For $\omega>\Omega+KR$,
$C(\omega)=C_{+}(\omega)>0$, $\dot{\psi}(\psi)$ is \emph{positive}
for any $\psi$, and the integral in Eq. (\ref{eq:C}) corresponds
to the time required for a D oscillator (with natural frequency $\omega$)
to complete a counterclockwise cycle over the unit circle, departing
from the initial phase $-\pi$ to the final one $+\pi$. Likewise,
for $\omega>\Omega-KR$, $C(\omega)=C_{-}(\omega)<0$, $\dot{\psi}(\psi)$
is \emph{negative} for any $\psi$, and the opposite of the integral
in Eq. (\ref{eq:C}) is the duration time of a clockwise cycle from
$+\pi$ to $-\pi$ . Therefore, the period of rotation over the unit
circle for a D oscillator with natural frequency $\omega$ is 
\begin{equation}
T(\omega)=\frac{1}{\left|C(\omega)\right|},
\end{equation}
and the absolute value of $C(\omega)$, $\left|C(\omega)\right|$,
is the number of cycles per time unit.

Now we turn our attention back to the $D$-oscillator phase density.
We can use Eqs. (\ref{eq:nd}), (\ref{eq:nd-cond-with-C}) and (\ref{eq:C+-})
to obtain 
\begin{equation}
n_{D}(\psi)=\intop_{-\infty}^{\Omega-KR}\frac{C_{-}(\omega)g(\omega)}{\dot{\psi}(\psi)}\,d\omega+\intop_{\Omega+KR}^{+\infty}\frac{C_{+}(\omega)g(\omega)}{\dot{\psi}(\psi)}\,d\omega.\label{eq:nd-C}
\end{equation}
By changing the integration variable from $\omega$ to $\chi=\omega-\Omega$
and replacing $\dot{\psi}(\psi)$ and $C_{\pm}(\omega)$ with their
respective expressions (See Eqs. (\ref{eq:D-f}) and (\ref{eq:C+-2})),
we can rewrite Eq. (\ref{eq:nd-C}) as 
\begin{equation}
n_{D}(\psi)=\frac{1}{2\pi}\intop_{\left|\chi\right|>KR}\frac{\chi g(\Omega+\chi)}{\chi-KR\sin\psi}\sqrt{1-\left(\frac{KR}{\chi}\right)^{2}}\,d\chi,\label{eq:nd-2}
\end{equation}
which is the final form of the phase density for D oscillators. Note
that, according to Eq. (\ref{eq:nd-2}), $n_{D}(\psi)\longrightarrow\frac{1}{2\pi}$
as $R\longrightarrow0$. Then, $n(\psi)=n_{S}(\psi)+n_{D}(\psi)\longrightarrow\frac{1}{2\pi}$
as $R\longrightarrow0$, since, as discussed before, $n_{S}(\psi)\longrightarrow0$
as $R\longrightarrow0^{+}$. This is consistent with the assumption
of a uniform phase density for the incoherent state. 

Using the formulas of the phase densities for S and D oscillators
(See Eqs. (\ref{eq:ns-1}) and (\ref{eq:nd-2})), one can obtain an
equation whose solution, for given $K$, $\alpha$ and $g$, consists
of $R$ and $\Omega$. To obtain such an equation, we first equate
the right-hand sides of Eqs. (\ref{eq:expected_oscillator_new_frame})
and (\ref{eq:mean-cn-new-frame-alpha}) with $n(\psi)$ defined by
Eq. (\ref{eq:n}). This gives 
\begin{equation}
Re^{i\alpha}=\intop_{-\pi}^{+\pi}e^{i\psi}\left[n_{S}(\psi)+n_{D}(\psi)\right]d\psi.\label{eq:Z-n}
\end{equation}

\emph{We are interested in non-trivial solutions, i.e. solutions with
non-zero $R$, meaning that synchronization occurs for the given parameters,
namely $K$, $\alpha$, and $g$. }So, $K$ is also finite. Otherwise,
all oscillators would be out of synchrony. By dividing both sides
of Eq. (\ref{eq:Z-n}) by the product $KR$ and using Eqs. (\ref{eq:ns-1})
and (\ref{eq:nd-2}), Eq. (\ref{eq:Z-n}) can be recast as

\begin{equation}
\frac{e^{i\alpha}}{K}=\intop_{-\frac{\pi}{2}}^{+\frac{\pi}{2}}e^{i\psi}g(\Omega+KR\sin\psi)\cos\psi\,d\psi+iJ,\label{eq:RO}
\end{equation}
where 
\begin{equation}
J=\frac{1}{2\pi iKR}\intop_{-\pi}^{+\pi}e^{i\psi}L(\psi)\,d\psi\label{eq:J-1}
\end{equation}
and 
\begin{equation}
L(\psi)=\intop_{KR}^{+\infty}\left[\frac{g(\Omega+\chi)}{\chi-KR\sin\psi}+\frac{g(\Omega-\chi)}{\chi+KR\sin\psi}\right]\chi\sqrt{1-\left(\frac{KR}{\chi}\right)^{2}}\,d\chi.\label{eq:L-1}
\end{equation}

We aim for a simpler form for Eq. (\ref{eq:J-1}). By changing the
integration variable in Eq. (\ref{eq:L-1}) from $\chi$ to $\gamma$,
defined by $\chi=KR\csc\gamma$ and $0<\gamma<\frac{\pi}{2}$, we
obtain 
\begin{equation}
L(\psi)=KR\intop_{0}^{+\frac{\pi}{2}}\left[\frac{h_{+}(\gamma)}{1-\sin\gamma\sin\psi}+\frac{h_{-}(\gamma)}{1+\sin\gamma\sin\psi}\right]\cot^{2}\gamma\,d\gamma,\label{eq:L-2}
\end{equation}
where 
\begin{equation}
h_{\pm}(\gamma)=g(\Omega\pm KR\csc\gamma)\label{eq:h}
\end{equation}

After substituting $L(\psi)$ in Eq. (\ref{eq:J-1}) by its definition
given in Eq. (\ref{eq:L-2}), a simple algebraic manipulation leads
to 
\begin{equation}
J=\frac{1}{2\pi i}\intop_{0}^{+\frac{\pi}{2}}\left\{ \left[h_{+}(\gamma)+h_{-}(\gamma)\right]I_{1}(\gamma)+\sin\gamma\left[h_{+}(\gamma)-h_{-}(\gamma)\right]I_{2}\left(\gamma\right)\right\} \cot^{2}\gamma\,d\gamma,\label{eq:J-2}
\end{equation}
where 
\begin{equation}
I_{1}(\gamma)=\intop_{-\pi}^{+\pi}e^{i\psi}f_{\gamma}^{(1)}(\psi)\,d\psi,\label{eq:I1}
\end{equation}
\begin{equation}
f_{\gamma}^{(1)}(\psi)=\frac{1}{1-\left(\sin\gamma\sin\psi\right)^{2}},
\end{equation}
\begin{equation}
I_{2}(\gamma)=\intop_{-\pi}^{+\pi}e^{i\psi}f_{\gamma}^{(2)}(\psi)\,d\psi,\label{eq:I2}
\end{equation}
and 
\begin{equation}
f_{\gamma}^{(2)}(\psi)=\frac{\sin\psi}{1-\left(\sin\gamma\sin\psi\right)^{2}}.
\end{equation}

Symmetry properties can be used to solve $I_{1}(\gamma)$ and $I_{2}(\gamma)$.
Eq. (\ref{eq:I1}) is the same as $I_{1}(\gamma)=\intop_{0}^{+\pi}\left[e^{i\psi}f_{\gamma}^{(1)}(\psi)+e^{i\left(\psi-\pi\right)}f_{\gamma}^{(1)}(\psi-\pi)\right]\,d\psi$.
Since $f_{\gamma}^{(1)}(\psi)=f_{\gamma}^{(1)}(\psi-\pi)$ and $e^{i\psi}=-e^{i\left(\psi-\pi\right)}$,
we conclude that $I_{1}(\gamma)=0$. Eq. (\ref{eq:I2}) can be written
as $I_{2}(\gamma)=\intop_{-\frac{\pi}{2}}^{+\frac{\pi}{2}}\left[e^{i\psi}f_{\gamma}^{(2)}(\psi)+e^{i\left(\pi-\psi\right)}f_{\gamma}^{(2)}(\pi-\psi)\right]\,d\psi$.
Considering that $f_{\gamma}^{(2)}(\psi)=f_{\gamma}^{(2)}(\pi-\psi)$
and $e^{i\psi}+e^{i\left(\pi-\psi\right)}=2i\sin\psi$, we have 
\begin{equation}
I_{2}(\gamma)=2i\intop_{-\frac{\pi}{2}}^{+\frac{\pi}{2}}\frac{\sin^{2}\psi}{1-\left(\sin\gamma\sin\psi\right)^{2}}\,d\psi,\label{eq:I2-form2}
\end{equation}
which has the alternative form 
\begin{equation}
I_{2}(\gamma)=\frac{2i}{\sin^{2}\gamma}\left[Q(\gamma)-\pi\right],\label{eq:I2-form3}
\end{equation}
where 
\begin{equation}
Q(\gamma)=\intop_{-\frac{\pi}{2}}^{+\frac{\pi}{2}}\frac{1}{1-\sin^{2}\gamma\sin{^{2}}\psi}\,d\psi.\label{eq:Q-1}
\end{equation}
Solving the integral in (\ref{eq:Q-1}) gives\footnote{According to Eq. (2.562-1) in Ref. \cite{Gradshteyn}, $\intop\frac{1}{a+b\sin{^{2}}x}\,dx=\frac{\text{sign}(a)}{\sqrt{a(a+b)}}\arctan\left(\sqrt{\frac{a+b}{a}}\tan x\right)$
for $\frac{b}{a}>-1$. If $a=1$ and $b=-\sin^{2}\gamma$, then, from
Eq. (\ref{eq:Q-1}), we have $Q(\gamma)=\lim_{\epsilon\to0^{+}}\left(\frac{1}{\sqrt{1-\sin^{2}\gamma}}\arctan\left(\sqrt{1-\sin^{2}\gamma}\tan\psi\right)\right|_{-\frac{\pi}{2}+\epsilon}^{+\frac{\pi}{2}-\epsilon}=\frac{\pi}{\left|\cos\gamma\right|}$.} 
\begin{equation}
Q(\gamma)=\frac{\pi}{\left|\cos\gamma\right|}.\label{eq:Q}
\end{equation}
Also, from Eqs. (\ref{eq:I2-form3}) and (\ref{eq:Q}),

\begin{equation}
I_{2}(\gamma)=\frac{2\pi i}{\sin^{2}\gamma}\left(\frac{1-\left|\cos\gamma\right|}{\left|\cos\gamma\right|}\right).\label{eq:I2-form4}
\end{equation}

Back to Eq. (\ref{eq:J-2}), we can eliminate the term with $I_{1}(\gamma)$
($I_{1}(\gamma)=0$, as mentioned above) and use Eqs. (\ref{eq:h})
and (\ref{eq:I2-form4}) to finally obtain

\begin{equation}
J=\intop_{0}^{+\frac{\pi}{2}}\left\{ g\left(\Omega+\frac{KR}{\sin\psi}\right)-g\left(\Omega-\frac{KR}{\sin\psi}\right)\right\} \frac{\cos\psi\left(1-\cos\psi\right)}{\sin^{3}\psi}\,d\psi.\label{eq:J}
\end{equation}

\emph{Eq. (\ref{eq:RO}), with $J$ defined by Eq. (\ref{eq:J}),
gives the values of $R$ and $\Omega$ for $K$ and $\alpha$ values
such that the KS model is in the synchronization state. }Eq. (\ref{eq:RO})
is equivalent to the system of equations 
\begin{align}
\intop_{-\frac{\pi}{2}}^{+\frac{\pi}{2}}g(\Omega+KR\sin\psi)\cos^{2}\psi\,d\psi= & \frac{1}{K}\cos\alpha\label{eq:ROsys-Re}\\
\intop_{-\frac{\pi}{2}}^{+\frac{\pi}{2}}g(\Omega+KR\sin\psi)\cos\psi\sin\psi\,d\psi+J= & \frac{1}{K}\sin\alpha\label{eq:ROsys-Im}
\end{align}

\begin{figure}
\begin{centering}
\subfloat[]{\includegraphics[width=0.5\linewidth]{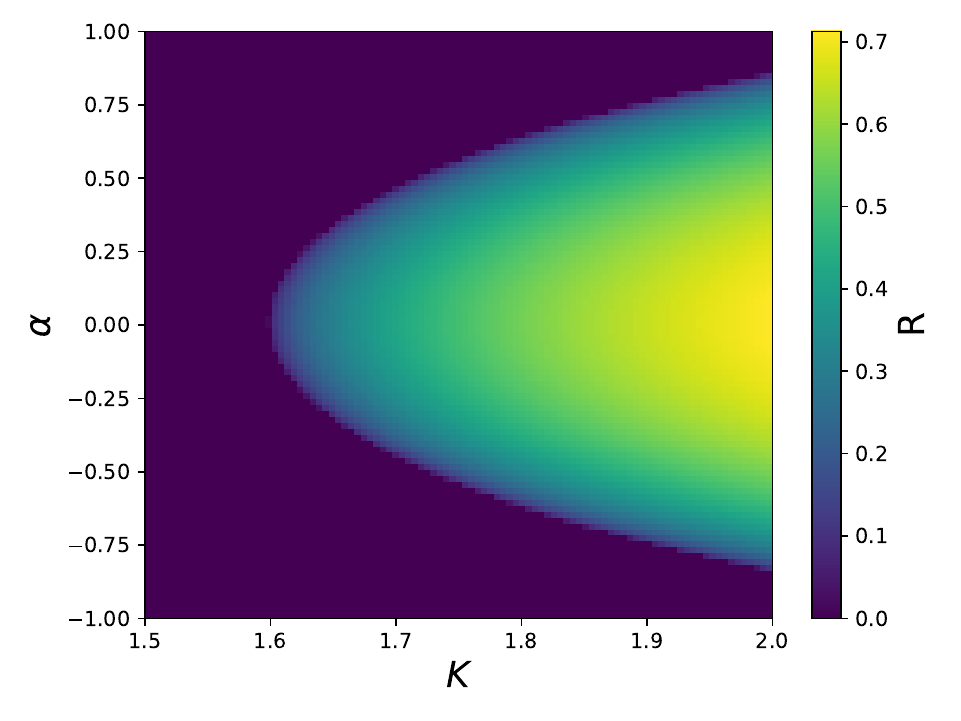}

}\subfloat[]{\includegraphics[width=0.5\linewidth]{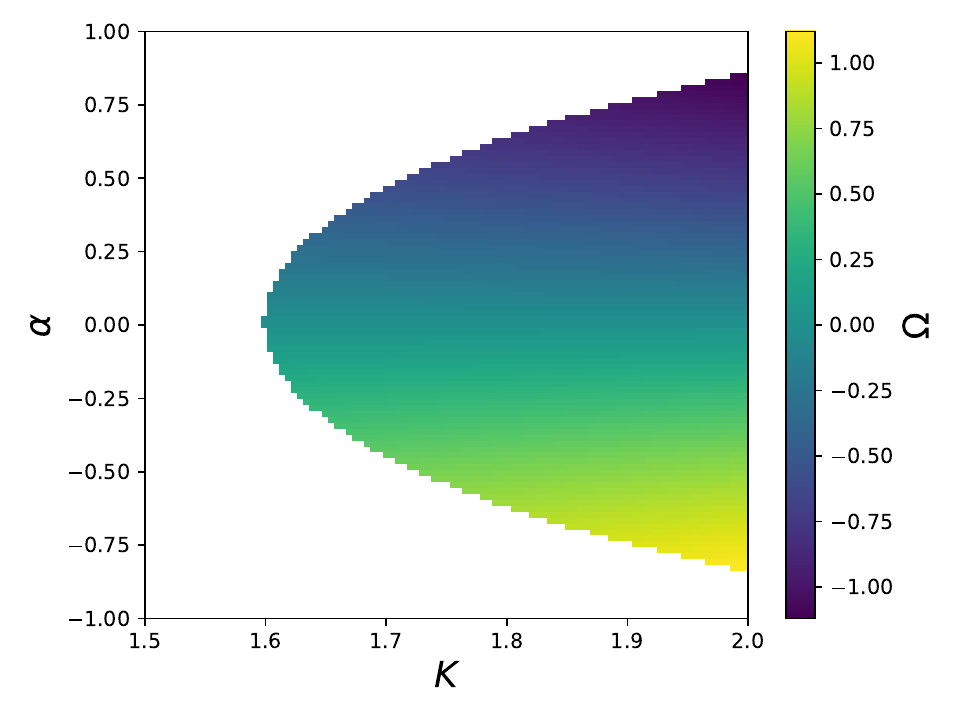}

}
\par\end{centering}
\begin{centering}
\subfloat[]{\includegraphics[width=0.5\linewidth]{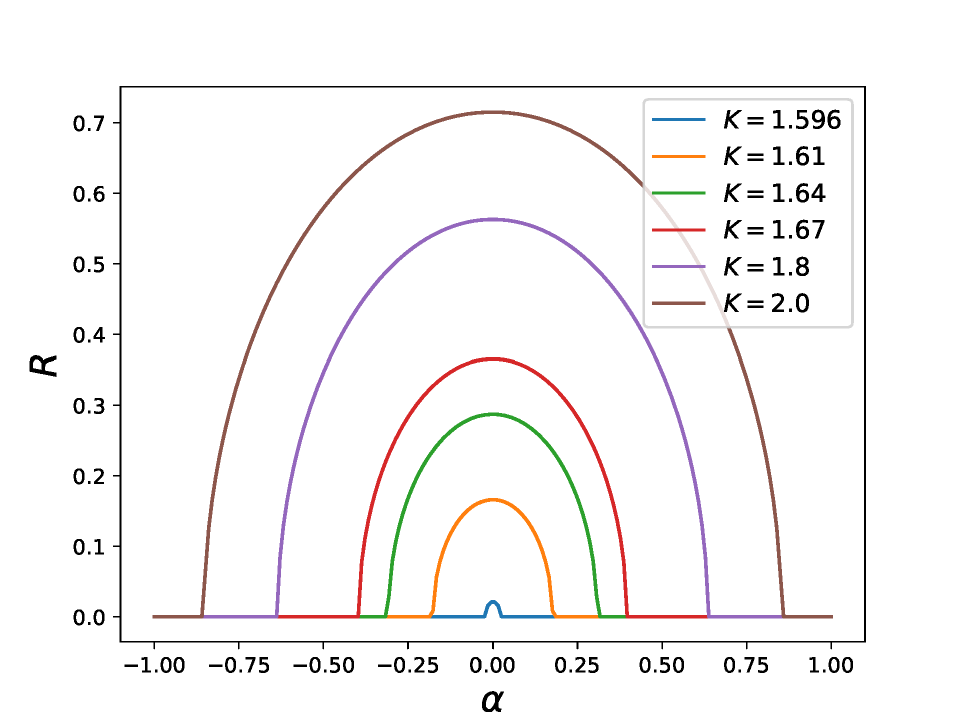}

}\subfloat[]{\includegraphics[width=0.5\linewidth]{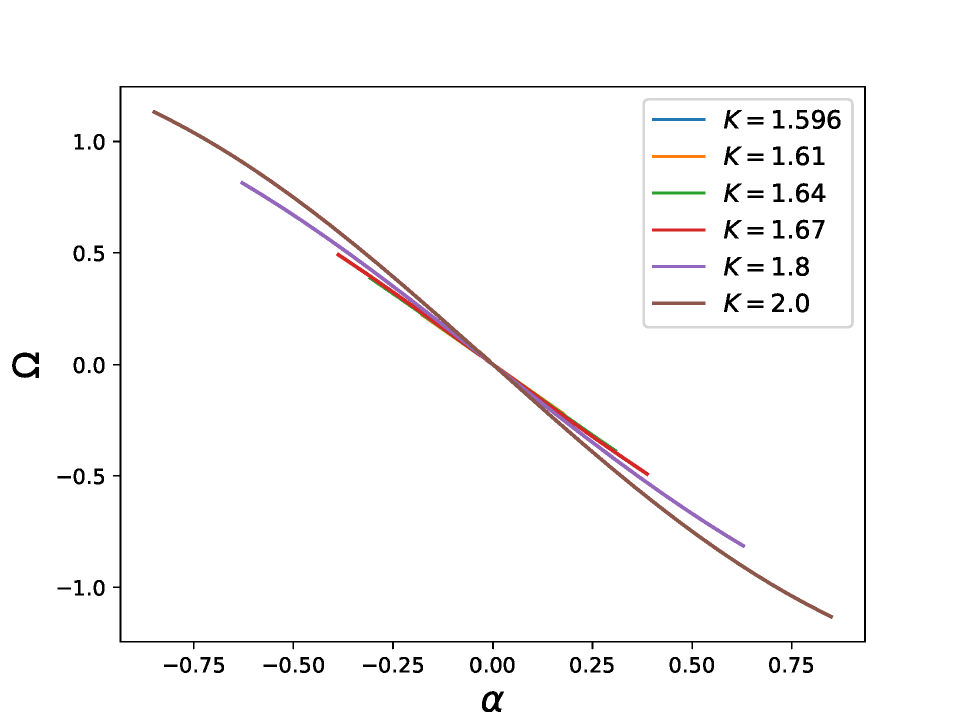}

}
\par\end{centering}
\caption{(a) Order-parameter diagram. (b) Synchronization-frequency diagram.
(c) Graphs of $R$ for constant $K$ and varying $\alpha$. (d) Graphs
of $\Omega$ for constant $K$ and varying $\alpha$. The values of
$R$ and $\Omega$ are numerical approximations to solutions of the
system of Eqs. (\ref{eq:ROsys-Re}-\ref{eq:ROsys-Im}). We assume
that $g$ is a standard normal density. }

\label{fig:ROGrid} 
\end{figure}

In Figs. \ref{fig:ROGrid}(a-d), we show numerical approximations
to solutions of the system (\ref{eq:ROsys-Re}-\ref{eq:ROsys-Im})
for different pairs of $K$ and $\alpha$ values. All numerical solutions
are computed by assuming that 
\begin{equation}
g(\omega)=\frac{1}{\sqrt{2\pi}}\exp\left(-\frac{\omega^{2}}{2}\right),\label{eq:normal-density}
\end{equation}
which is the standard normal density, and using the software library
MINPACK\cite{More}. If convergence is not achieved, we assume that
the system (\ref{eq:ROsys-Re}-\ref{eq:ROsys-Im}) has no solution,
set $R=0$, and attribute no specific value to $\Omega$. This means
that the KS model is in the incoherent state.

Figure\ref{fig:ROGrid}(a) shows a $100\times100$ resolution grid
of points $(K,\alpha)$. Each point has a color defined according
to the value of $R$, e.g. dark blue for $R=0$. A well-defined boundary
separates two regions, one with $R=0$ and another where $R>0$. Figure
\ref{fig:ROGrid}(a) can be seen as a \emph{phase diagram} pointing
out the transition from a fully desynchronized state to a hybrid state
consisting of both D and S oscillators.

Figure\ref{fig:ROGrid}(b) shows a similar grid to that of Fig.\ref{fig:ROGrid}(a).
The grid has the same set of points $(K,\alpha)$, but the color of
each point is defined by the corresponding value of $\Omega$. If
$\Omega$ is not defined for a specific point, we associate this point
with the white color.

In Figs. \ref{fig:ROGrid}(c) and (d), we show slices of the three-dimensional
graphs of Figs. \ref{fig:ROGrid}(a) and (b). For each two-dimensional
profile, $K$ is kept fixed, and $\alpha$ varies between $-1$ and
$+1$. The graphs of Figure \ref{fig:ROGrid}(c) suggest that, for
constant $K$, $R$ is an even function of $\alpha$ with a maximum
at $\alpha=0$. In Figure \ref{fig:ROGrid}(d), the graphs indicate
that $\Omega$ varies monotonically as an odd function of $\alpha$.
Moreover, at least for the set of $K$ values considered, $\Omega$
is more sensitive to variations in $\alpha$ than in $K$.

\section{Density of instantaneous frequencies \label{sec:G}}

In the previous section, we showed how to obtain known results from
the KS theory relevant to this work. We are now able to proceed towards
our core result: the density of instantaneous frequencies in the KS
model, which we represent by the probability density function $G$.
The quantity $G(\nu)\,d\nu$ is then the probability that a KS oscillator
$i$ has its fixed-frame instantaneous frequency, $\dot{\theta}_{i},$
in the interval $\left[\nu,\,\nu+d\nu\right)$.

\emph{In the process of obtaining $G$, we are concerned about the
case $R>0$ (synchronization state).} According to Eqs. (\ref{eq:psi-dot-def})
and (\ref{eq:KS-KR}), for $R=0$, $\dot{\theta}_{i}=\omega_{i}$.
Then, in the incoherent state, the densities of instantaneous and
natural frequencies are identical, i.e. $G(\nu)=g(\nu)$. 

Before we show how to obtain the density $G$, we introduce some basic
facts and definitions to simplify the notation. First, the product
$KR$ is defined by 
\begin{equation}
a\equiv KR,\label{eq:a}
\end{equation}
and 
\begin{equation}
\widetilde{x}\equiv\frac{x-\Omega}{a}\label{eq:tilde-def}
\end{equation}
for a generic variable $x$. We can then rewrite Eqs. (\ref{eq:attractor-1}),
(\ref{eq:D-f}), and (\ref{eq:C+-}) as 
\begin{equation}
\psi^{*}(\omega)=\sin^{-1}\left(\widetilde{\omega}\right),\label{eq:attractor-2}
\end{equation}
\begin{equation}
\dot{\psi}(\psi)=a\left(\widetilde{\omega}-\sin\psi\right),\label{eq:D-f-2}
\end{equation}
and 
\begin{equation}
C\left(\omega\right)=\frac{a}{2\pi}\sgn{\widetilde{\omega}}\sqrt{\widetilde{\omega}^{2}-1}.\label{eq:C+-3}
\end{equation}
In Eq. (\ref{eq:C+-3}), $\sgn{\widetilde{\omega}}$ is the sign of
$\widetilde{\omega}$ and $\left|\widetilde{\omega}\right|>1$. Another
set of useful definitions is 
\begin{equation}
\dot{\psi}\equiv\nu-\Omega,\label{eq: nu}
\end{equation}
\begin{equation}
\Delta\nu\equiv\widetilde{\nu}-\widetilde{\omega},\label{eq:Delta-nu}
\end{equation}
and the function 
\begin{equation}
f_{\nu}(\psi)\equiv\Delta\nu+\sin(\psi).\label{eq:fnu}
\end{equation}
The symbol $\dot{\psi}$ is used to denote the rotating-frame instantaneous
frequency of an oscillator with a fixed-frame instantaneous frequency
$\nu$. Definitions (\ref{eq: nu}) and (\ref{eq:Delta-nu}) imply
\begin{equation}
\widetilde{\nu}=\frac{\dot{\psi}}{a}\label{eq:nu-tilde}
\end{equation}
and 
\begin{equation}
\Delta\nu=\frac{\dot{\psi}}{a}-\widetilde{\omega}.\label{eq:Delta-nu-2}
\end{equation}

From the formulas for $n_{S}(\psi|\omega)$ and $n_{D}(\psi|\omega)$,
given by Eqs. (\ref{eq:ns-cond}) and (\ref{eq:nd-cond-with-C}),
the phase density for an oscillator with known natural frequency $\omega$
is
\begin{equation}
n(\psi|\omega)=\begin{cases}
\delta\left[\psi-\psi^{*}\left(\omega\right)\right],\quad & \left|\widetilde{\omega}\right|\leq1\\
\\
\frac{C(\omega)}{\dot{\psi}(\psi)}, & \left|\widetilde{\omega}\right|>1
\end{cases}\label{eq:n-cond}
\end{equation}
where $\psi^{*}\left(\omega\right)$ and $C\left(\omega\right)$ is
defined by Eqs. (\ref{eq:attractor-2}) and (\ref{eq:C+-3}).

We now have enough mathematical tools to determine $G$. As a first
step, we define $p(\dot{\psi}|\omega)$ as the probability density
that an oscillator has a rotating-frame instantaneous frequency $\dot{\psi}$
given that the oscillator's natural frequency is $\omega$. From the
random variable transformation theorem \cite{Gillepsie83}, we have
\begin{equation}
p(\dot{\psi}|\omega)=\int_{-\pi}^{+\pi}\delta[\dot{\psi}-\dot{\psi}(\psi)]n(\psi|\omega)\,d\psi\label{eq:p-cond}
\end{equation}
In Eq. (\ref{eq:p-cond}), $\dot{\psi}$ is the argument of function
$p(\dot{\psi}|\omega)$, $\dot{\psi}(\psi)$ is the function of $\psi$
defined by Eq. (\ref{eq:D-f-2}), and $n(\psi|\omega)$ is the conditional
density (\ref{eq:n-cond}).

From Eqs. (\ref{eq:n-cond}) and(\ref{eq:p-cond}), 
\begin{equation}
p(\dot{\psi}|\omega)=\begin{cases}
p_{S}(\dot{\psi}|\omega),\quad & \left|\widetilde{\omega}\right|\leq1\\
\\
p_{D}(\dot{\psi}|\omega), & \left|\widetilde{\omega}\right|>1
\end{cases}\label{eq:p-cond-1}
\end{equation}
where 
\begin{equation}
p_{S}(\dot{\psi}|\omega)=\frac{1}{a}\int_{-\pi}^{+\pi}\delta\left[\psi-\psi^{*}\left(\omega\right)\right]\delta\left[f_{\nu}(\psi)\right]\,d\psi\label{eq:ps-cond}
\end{equation}
and 
\begin{equation}
p_{D}(\dot{\psi}|\omega)=\frac{1}{2\pi a}\sgn{\widetilde{\omega}}\sqrt{\widetilde{\omega}^{2}-1}\int_{-\pi}^{+\pi}\frac{\delta\left[f_{\nu}(\psi)\right]}{\widetilde{\omega}-\sin\psi}\,d\psi.\label{eq:pd-cond}
\end{equation}

By solving the integral in Eq. (\ref{eq:ps-cond}), we get $p_{S}(\dot{\psi}|\omega)=\delta\left\{ af_{\nu}\left[\psi^{*}\left(\omega\right)\right]\right\} $.
So, from (\ref{eq:attractor-2}), (\ref{eq: nu}), (\ref{eq:Delta-nu}),
and (\ref{eq:fnu}), 
\begin{equation}
p_{S}(\dot{\psi}|\omega)=\delta(\dot{\psi}),\label{eq:ps-cond-1}
\end{equation}
which expresses the certainty that S oscillators have rotating-frame
instantaneous frequencies equal to zero, i.e., their fixed-frame instantaneous
frequencies are equal to $\Omega$.

To solve the integral in Eq. (\ref{eq:pd-cond}), we write $\delta\left[f_{\nu}(\psi)\right]$
as 
\begin{equation}
\delta\left[f_{\nu}(\psi)\right]=\sum_{\beta\in O\left[f_{\nu}\right]}\frac{\delta(\psi-\beta)}{\left|f_{\nu}'(\beta)\right|},\label{eq:delta-fnu}
\end{equation}
where $f_{\nu}'$ is the derivative of $f_{\nu}$, and $\beta$ runs
through $O\left[f_{\nu}\right]$, defined as the set of the \emph{simple
zeros} of $f_{\nu}$. If $O\left[f_{\nu}\right]$ is an empty set,
which is the case for $\left|\Delta\nu\right|\geq1$, $\delta\left[f_{\nu}(\psi)\right]=0$.
For $\left|\Delta\nu\right|<1$, $f_{\nu}$ has two simple zeros:
$\beta_{1}=\sin^{-1}\left(-\Delta\nu\right)$ and $\beta_{2}=\pi-\beta_{1}$.
Since $\left|f_{\nu}'\left(\beta_{1,2}\right)\right|=\left|\cos\left[\sin^{-1}\left(\Delta\nu\right)\right]\right|=\sqrt{1-\left(\Delta\nu\right)^{2}}$,
Eq. (\ref{eq:delta-fnu}) gives 
\begin{equation}
\delta\left[f_{\nu}(\psi)\right]=\begin{cases}
\frac{1}{\sqrt{1-\left(\Delta\nu\right)^{2}}}\left[\delta\left(\psi-\beta_{1}\right)+\delta\left(\psi-\beta_{2}\right)\right],\quad & \left|\Delta\nu\right|<1\\
\\
0, & \left|\Delta\nu\right|\geq1.
\end{cases}\label{eq:delta-fnu-1}
\end{equation}

From Eqs. (\ref{eq:pd-cond}) and (\ref{eq:delta-fnu-1}), we have
\begin{equation}
p_{D}(\dot{\psi}|\omega)=\begin{cases}
\frac{\sgn{\widetilde{\omega}}}{\pi a\widetilde{\nu}}\sqrt{\frac{\widetilde{\omega}^{2}-1}{1-\left(\Delta\nu\right)^{2}}},\quad & \left|\Delta\nu\right|<1\\
\\
0, & \left|\Delta\nu\right|\geq1,
\end{cases}\label{eq:pd-cond-1}
\end{equation}
whose explicit dependence on $\dot{\psi}$ can be obtained using Eqs.
(\ref{eq:nu-tilde}) and (\ref{eq:Delta-nu-2}).

With Eqs. (\ref{eq:ps-cond-1}) and (\ref{eq:pd-cond-1}), we complete
the definition of the conditional density $p(\dot{\psi}|\omega)$,
given by (\ref{eq:p-cond-1}). In a similar way to what we have done
to obtain the density of phases (See Eqs. (\ref{eq:n}), (\ref{eq:ns})
and (\ref{eq:nd})), we can define the densities of instantaneous
frequencies as 
\begin{equation}
p(\dot{\psi})=p_{S}(\dot{\psi})+p_{D}(\dot{\psi}),\label{eq:p}
\end{equation}
where 
\begin{equation}
p_{S}(\dot{\psi})=\intop_{\left|\widetilde{\omega}\right|\leq1}p_{S}(\dot{\psi}|\omega)g(\omega)\,d\omega\label{eq:ps}
\end{equation}
and 
\begin{equation}
p_{D}(\dot{\psi})=\intop_{\left|\widetilde{\omega}\right|>1}p_{D}(\dot{\psi}|\omega)g(\omega)\,d\omega.\label{eq:pd}
\end{equation}
In Eqs. (\ref{eq:ps}) and (\ref{eq:pd}), $p_{S}(\dot{\psi})$ and
$p_{D}(\dot{\psi})$ are the densities of instantaneous frequencies
in the rotating frame for S and D oscillators, respectively.

Since the integration domain in (\ref{eq:ps}) is the interval $\left[\Omega-a,\Omega+a\right]$,
substituting (\ref{eq:ps-cond-1}) in (\ref{eq:ps}) results in 
\begin{equation}
p_{S}(\dot{\psi})=S(K,\alpha)\delta(\dot{\psi}),\label{eq:ps-final}
\end{equation}
where 
\begin{equation}
S(K,\alpha)=\int_{\Omega-a}^{\Omega+a}g(\omega)d\omega.\label{eq:S}
\end{equation}
The quantity $S(K,\alpha)$ has an important meaning: $S(K,\alpha)$
is the probability that an oscillator is synchronized. So, $S(K,\alpha)$
quantifies the fraction of S oscillators. Note that $S(K,\alpha)$
has implicit dependencies on $K$ and $\alpha$ through $\Omega$
and $a$. As already mentioned, $a=KR$, and the pair $\left\{ \Omega,R\right\} $
is the solution of the system of Eqs. (\ref{eq:ROsys-Re}-\ref{eq:ROsys-Im}).

In order to obtain a formula for $p_{D}(\dot{\psi})$ from Eq. (\ref{eq:pd}),
it is useful to define $D(\nu)$ as the set of real numbers $\widetilde{\omega}$
such that $\left|\widetilde{\omega}\right|>1$ and $\left|\Delta\nu\right|<1$.
That is, 
\begin{equation}
D(\nu)\equiv\left\{ \widetilde{\omega}\in{\rm I\!R}-\left[-1,+1\right]\mid\widetilde{\nu}-1<\widetilde{\omega}<\widetilde{\nu}+1\right\} .\label{eq:D-nu}
\end{equation}
Note that, for $\widetilde{\nu}=0$, $D(\nu)$ is the empty set. For
finite $\widetilde{\nu}$, $D(\nu)$ is an open interval $\left(\widetilde{\omega}_{\nu}^{-},\widetilde{\omega}_{\nu}^{+}\right)$,
whose endpoints $\widetilde{\omega}_{\nu}^{-}$ and $\widetilde{\omega}_{\nu}^{+}$
change according to the value of $\nu$. The endpoints of $D(\nu)$
are given in Table \ref{tab:T1} for different intervals of $\tilde{\nu}$.
\begin{table}
\begin{centering}
\begin{tabular}{ccccc}
\toprule 
 & $\tilde{\nu}\leq-2$  & $-2<\tilde{\nu}<0$  & $0<\tilde{\nu}<+2$  & $+2\leq\tilde{\nu}$\tabularnewline
\midrule 
$\widetilde{\omega}_{\nu}^{+}$  & {\small{}{}{}$\tilde{\nu}+1$}  & {\small{}{}{}$-1$}  & {\small{}{}{}$\tilde{\nu}+1$}  & {\small{}{}{}$\tilde{\nu}+1$}\tabularnewline
$\widetilde{\omega}_{\nu}^{-}$  & {\small{}{}{}$\tilde{\nu}-1$}  & {\small{}{}{}$\tilde{\nu}-1$}  & {\small{}{}{}$+1$}  & {\small{}{}{}$\tilde{\nu}-1$}\tabularnewline
\bottomrule
\end{tabular}
\par\end{centering}
\caption{Endpoints of the open interval $D(\nu)=\left(\widetilde{\omega}_{\nu}^{-},\widetilde{\omega}_{\nu}^{+}\right)$,
defined by (\ref{eq:D-nu}). For $\widetilde{\nu}=0$, $D(\nu)$ is
the empty set.}
\label{tab:T1} 
\end{table}

A more compact way of defining $\widetilde{\omega}_{\nu}^{-}$ and
$\widetilde{\omega}_{\nu}^{+}$ is 
\begin{equation}
\widetilde{\omega}_{\nu}^{\pm}=\left(\tilde{\nu}\pm2\right)\Theta\left[\tilde{\nu}\left(\tilde{\nu}\pm2\right)\right]\mp1,\label{eq:ends-omega}
\end{equation}
where $\Theta$ denotes the Heaviside step function with the standard
definition $\Theta(0)=\frac{1}{2}$.

Let $\overline{D}(\nu)$ be the set of real numbers $\widetilde{\omega}$
such that $\left|\widetilde{\omega}\right|>1$ and $\left|\Delta\nu\right|\geq1$.
So, from (\ref{eq:pd-cond-1}) and (\ref{eq:pd}), 
\begin{equation}
p_{D}(\dot{\psi})=\int_{D(\nu)}p_{D}(\dot{\psi}|\omega)g(\omega)\,d\omega+\int_{\overline{D}(\nu)}p_{D}(\dot{\psi}|\omega)g(\omega)\,d\omega.\label{eq:pd-2}
\end{equation}
Since $p_{D}(\dot{\psi}|\omega)=0$ for $\left|\Delta\nu\right|\geq1$
(See Eq. (\ref{eq:pd-cond-1})), the second integral in (\ref{eq:pd-2})
is zero. The first integral is also zero only if $D(\nu)$ is empty,
that is, if $\widetilde{\nu}=0$ ($\dot{\psi}=0$). If $\widetilde{\nu}\neq0$
($\dot{\psi}\neq0$), the first integral can be determined using the
endpoints of $D(\nu)$ as integration limits and the expression of
$p_{D}(\dot{\psi}|\omega)$ for $\left|\Delta\nu\right|<1$ defined
in Eq. (\ref{eq:pd-cond-1}). Thus, Eq. (\ref{eq:pd-2}) can be rewritten
as 
\begin{equation}
p_{D}(\dot{\psi})=\frac{a}{\pi\dot{\psi}}\int_{\widetilde{\omega}_{\dot{\psi}+\Omega}^{-}}^{\widetilde{\omega}_{\dot{\psi}+\Omega}^{+}}\sgn{\widetilde{\omega}}\sqrt{\frac{\widetilde{\omega}^{2}-1}{1-\left(\frac{\dot{\psi}}{a}-\widetilde{\omega}\right)^{2}}}g(a\widetilde{\omega}+\Omega)\,d\widetilde{\omega}\label{eq:pd-final}
\end{equation}
for non-zero values of $\dot{\psi}$, and $p_{D}(\dot{\psi})=0$ for
$\dot{\psi}=0$. From (\ref{eq:ends-omega}), the limits of integration
in (\ref{eq:pd-final}) are 
\begin{equation}
\widetilde{\omega}_{\dot{\psi}+\Omega}^{\pm}=\left(\frac{\dot{\psi}}{a}\pm2\right)\Theta\left[\frac{\dot{\psi}}{a}\left(\frac{\dot{\psi}}{a}\pm2\right)\right]\mp1.
\end{equation}
In order to show the explicit dependence of $p_{D}(\dot{\psi})$ on
$\dot{\psi}$, we used (\ref{eq: nu}), (\ref{eq:nu-tilde}), and
(\ref{eq:Delta-nu-2}). We also changed the previous integration variable,
$\omega$, to $\widetilde{\omega}$.

Equations (\ref{eq:p}), (\ref{eq:ps-final}) and (\ref{eq:pd-final})
give a full description of the density of instantaneous frequencies
in the rotating frame. Our goal now is to obtain the density of fixed-frame
instantaneous frequencies, which we defined, at the beginning of this
section, as a probability density function $G(\nu).$ 

Instantaneous frequencies in the rotating and fixed frames are related
through Eq. (\ref{eq: nu}). Therefore, $G(\nu)=p(\nu-\Omega)$, which
is the same as 
\begin{equation}
G(\nu)=p_{S}(\nu-\Omega)+p_{D}(\nu-\Omega).\label{eq:G-p-1}
\end{equation}
Let $G_{S}(\nu)=p_{S}(\nu-\Omega)$ and $G_{D}(\nu)=p_{D}(\nu-\Omega)$.
From (\ref{eq:ps-final}), (\ref{eq:pd-final}), and (\ref{eq:G-p-1}),
we have 
\begin{equation}
G(\nu)=G_{S}(\nu)+G_{D}(\nu),\label{eq:G}
\end{equation}
where 
\begin{equation}
G_{S}(\nu)=S(K,\alpha)\delta(\nu-\Omega),\label{eq:Gs}
\end{equation}
\begin{equation}
G_{D}(\nu)=\frac{1}{\pi\left|\widetilde{\nu}\right|}\int_{\widetilde{\omega}_{\nu}^{-}}^{\widetilde{\omega}_{\nu}^{+}}\sqrt{\frac{\widetilde{\omega}^{2}-1}{1-(\widetilde{\nu}-\widetilde{\omega})^{2}}}g(a\widetilde{\omega}+\Omega)\,d\widetilde{\omega},\label{eq:Gd}
\end{equation}
for $\left|\nu-\Omega\right|>0$, and $G_{D}(\Omega)=0$. Note that
we replaced $\frac{\sgn{\widetilde{\omega}}}{\widetilde{\nu}}$ with
$\frac{1}{\left|\widetilde{\nu}\right|}$ to obtain Eq. (\ref{eq:Gd}).
This is possible due to the integral-limits signs, which can be determined
from Table \ref{tab:T1}. If $\widetilde{\nu}$ is positive, both
$\widetilde{\omega}_{\nu}^{-}$ and $\widetilde{\omega}_{\nu}^{+}$
are positive. If $\widetilde{\nu}$ is negative, both $\widetilde{\omega}_{\nu}^{-}$
and $\widetilde{\omega}_{\nu}^{+}$ are negative. So, the integration
variable, $\widetilde{\omega}$, which is in the interval $\widetilde{\omega}_{\nu}^{-}<\widetilde{\omega}<\widetilde{\omega}_{\nu}^{+}$,
has the same sign as $\widetilde{\nu}$. This implies $\widetilde{\nu}=\sgn{\widetilde{\omega}}\left|\widetilde{\nu}\right|$.

An alternative formula for Eq. (\ref{eq:Gd}) can be obtained by changing
the integration variable from $\widetilde{\omega}$ to $\psi=\sin^{-1}(\widetilde{\omega}-\widetilde{\nu})$.
This change results in: $\widetilde{\omega}^{2}-1=(\sin\psi+\widetilde{\nu})^{2}-1$,
$\sqrt{1-(\widetilde{\nu}-\widetilde{\omega})^{2}}=\cos\psi$, and
$g(a\widetilde{\omega}+\Omega)=g(a\sin\psi+\nu)$. In addition, if
$\psi_{\nu}^{-}$ and $\psi_{\nu}^{+}$ denote the new integral limits,
then, from $\psi_{\nu}^{\pm}=\sin^{-1}(\widetilde{\omega}_{\nu}^{\pm}-\widetilde{\nu})$
and $\Theta(x)+\Theta(-x)=1$, 
\begin{equation}
\psi_{\nu}^{\pm}=\sin^{-1}\left\{ -\left(\tilde{\nu}\pm2\right)\Theta\left[-\tilde{\nu}\left(\tilde{\nu}\pm2\right)\right]\pm1\right\} .\label{eq:ends-psi}
\end{equation}
The formulas above allow us to rewrite Eq. (\ref{eq:Gd}) as 
\begin{equation}
G_{D}(\nu)=\frac{1}{\pi\left|\widetilde{\nu}\right|}\int_{\psi_{\nu}^{-}}^{\psi_{\nu}^{+}}\sqrt{(\sin\psi+\widetilde{\nu})^{2}-1}g(a\sin\psi+\nu)\,d\psi.\label{eq:Gd-1}
\end{equation}

Equations (\ref{eq:G}), (\ref{eq:Gs}), and (\ref{eq:Gd-1}) are
identical to those which define the density of instantaneous frequencies
in the Kuramoto model \cite{Fonseca20}. However, although the formulas
in both models depend on $R$ and $\Omega$ in the same manner, $R$
and $\Omega$ depend implicitly on $\alpha$ in the KS model, while
the parameter $\alpha$ is not defined in the Kuramoto model.

We conclude this section in a summary fashion presenting again our
main result with the original notation restored and together with
all auxiliary equations. From Eqs. (\ref{eq:G}), (\ref{eq:Gs}) and
(\ref{eq:Gd}), 
\begin{equation}
G(\nu)=S(K,\alpha)\delta(\nu-\Omega)+G_{D}(\nu),\label{eq:G-1}
\end{equation}
where 
\begin{equation}
S(K,\alpha)=\int_{\Omega-KR}^{\Omega+KR}g(\omega)d\omega,\label{eq:S-1}
\end{equation}
and 
\begin{equation}
G_{D}(\nu)=\frac{KR}{\pi\left[\left|\nu-\Omega\right|+\Theta(-\left|\nu-\Omega\right|)\right]}\int_{\widetilde{\omega}_{\nu}^{-}}^{\widetilde{\omega}_{\nu}^{+}}\sqrt{\frac{\widetilde{\omega}^{2}-1}{1-\left(\frac{\nu-\Omega}{KR}-\widetilde{\omega}\right)^{2}}}g(\Omega+KR\widetilde{\omega})\,d\widetilde{\omega},\label{eq:Gd-2}
\end{equation}
which have the integral limits

\begin{equation}
\widetilde{\omega}_{\nu}^{\pm}=\left(\frac{\nu-\Omega}{KR}\pm2\right)\Theta\left[\frac{\nu-\Omega}{KR}\left(\frac{\nu-\Omega}{KR}\pm2\right)\right]\mp1.\label{eq:ends-omega-1}
\end{equation}
We introduced the function $\Theta(-\left|\nu-\Omega\right|)$ in
(\ref{eq:Gd-2}) so that $G_{D}(\Omega)=0$, as stated earlier. The
quantities $R$ and $\Omega$, which appear in Eqs. (\ref{eq:G-1}),
(\ref{eq:S-1}), (\ref{eq:Gd-2}), and (\ref{eq:ends-omega-1}), form
the solution of Eq. (\ref{eq:RO}), namely 
\begin{equation}
\frac{e^{i\alpha}}{K}=\intop_{-\frac{\pi}{2}}^{+\frac{\pi}{2}}e^{i\psi}g(\Omega+KR\sin\psi)\cos\psi\,d\psi+iJ,\label{eq:RO-1}
\end{equation}
where 
\begin{equation}
J=\intop_{0}^{+\frac{\pi}{2}}\left\{ g\left(\Omega+\frac{KR}{\sin\psi}\right)-g\left(\Omega-\frac{KR}{\sin\psi}\right)\right\} \frac{\cos\psi\left(1-\cos\psi\right)}{\sin^{3}\psi}\,d\psi.\label{eq:J-3}
\end{equation}
Interestingly, from Eqs. (\ref{eq:G-1}) to (\ref{eq:J-3}), Eq. (\ref{eq:RO-1})
is the only one in which the parameter $\alpha$ is present. 

\section{Application: Gaussian density of natural frequencies \label{sec:Application}}

\begin{figure}
\begin{centering}
\subfloat[]{\includegraphics[width=0.45\linewidth]{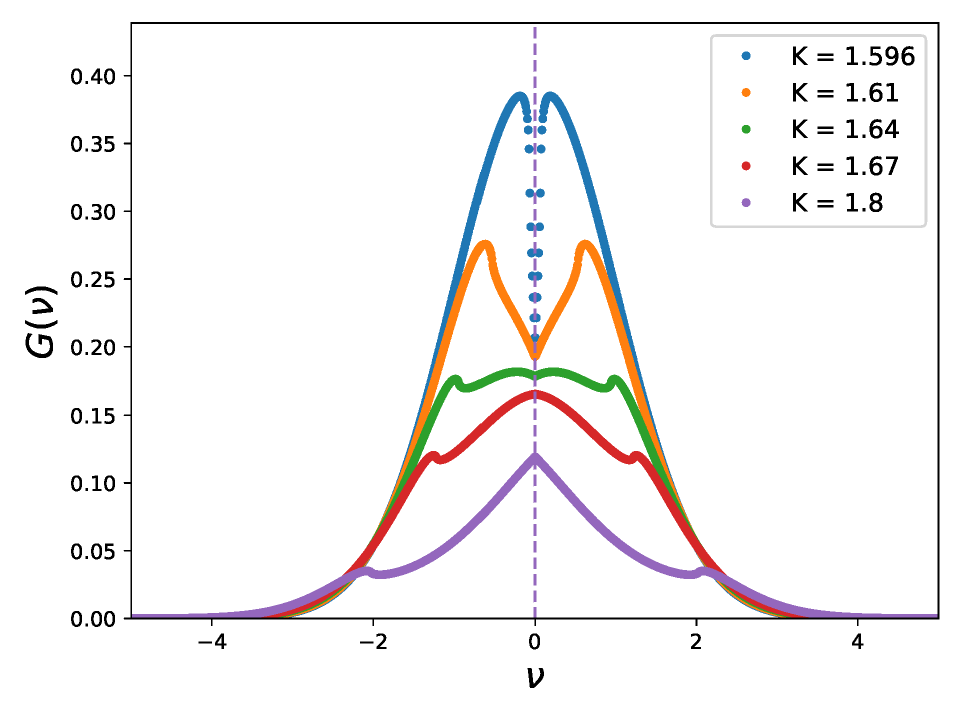}

}\subfloat[]{\begin{centering}
\includegraphics[width=0.45\linewidth]{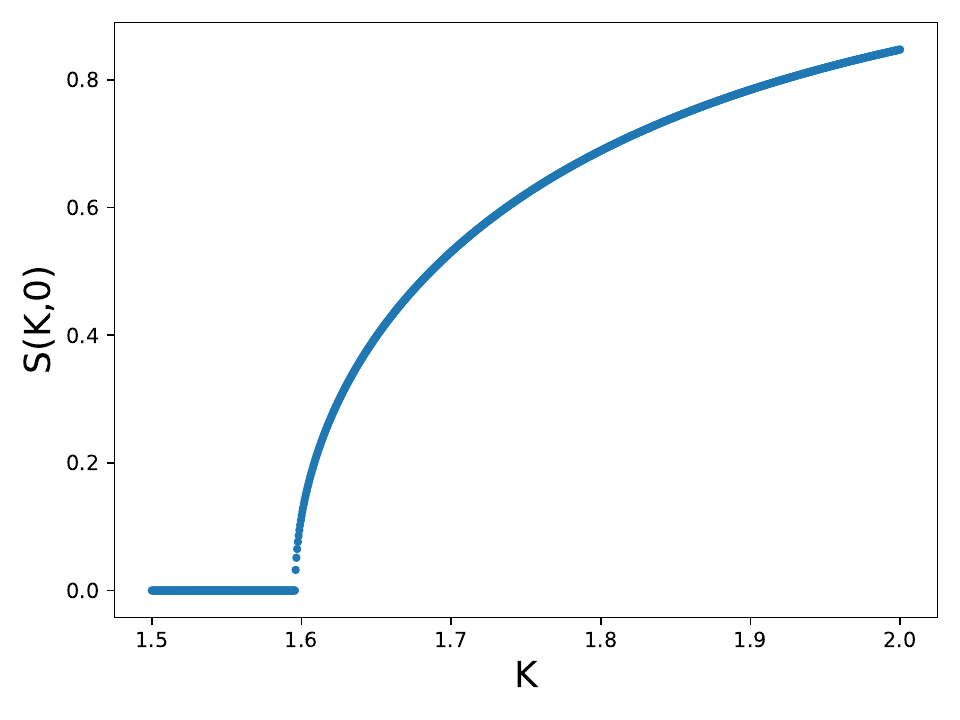} 
\par\end{centering}
}
\par\end{centering}
\caption{(a) Graphs of $G_{D}$ for $\alpha=0$. The dashed vertical line represents
the delta term in (\ref{eq:G-1}). (b) Fraction of synchronized oscillators,
$S(K,\alpha)$, for $\alpha=0$.}
\label{fig:G-Kur-Gauss} 
\end{figure}

\begin{figure}
\begin{centering}
\subfloat[]{\includegraphics[width=0.45\linewidth]{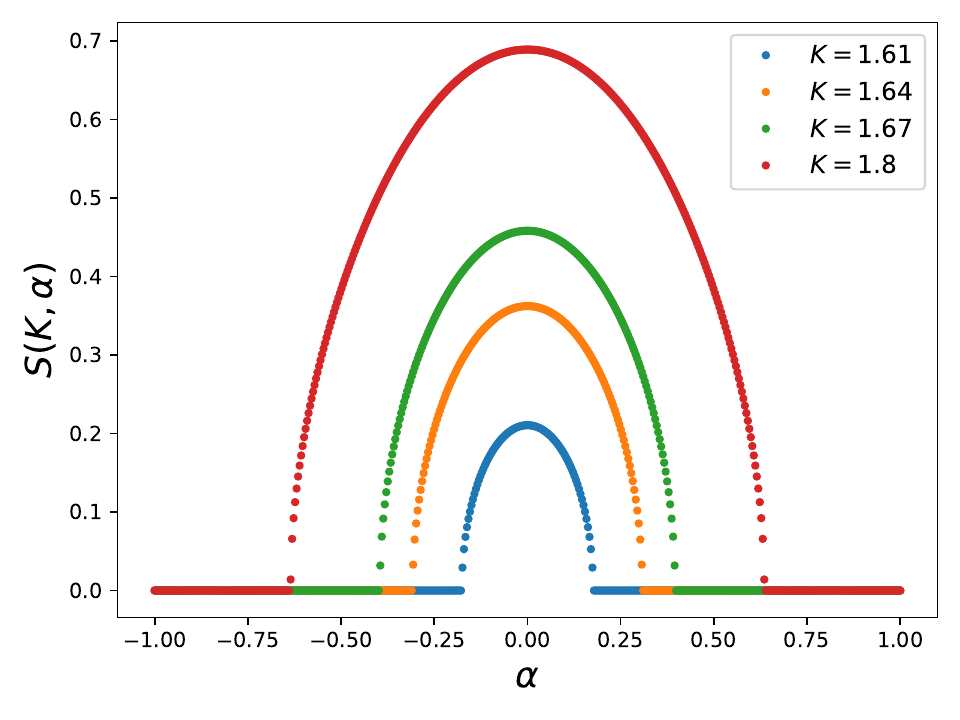}

}
\par\end{centering}
\begin{centering}
\subfloat[$K=1.61$]{\includegraphics[width=0.45\linewidth]{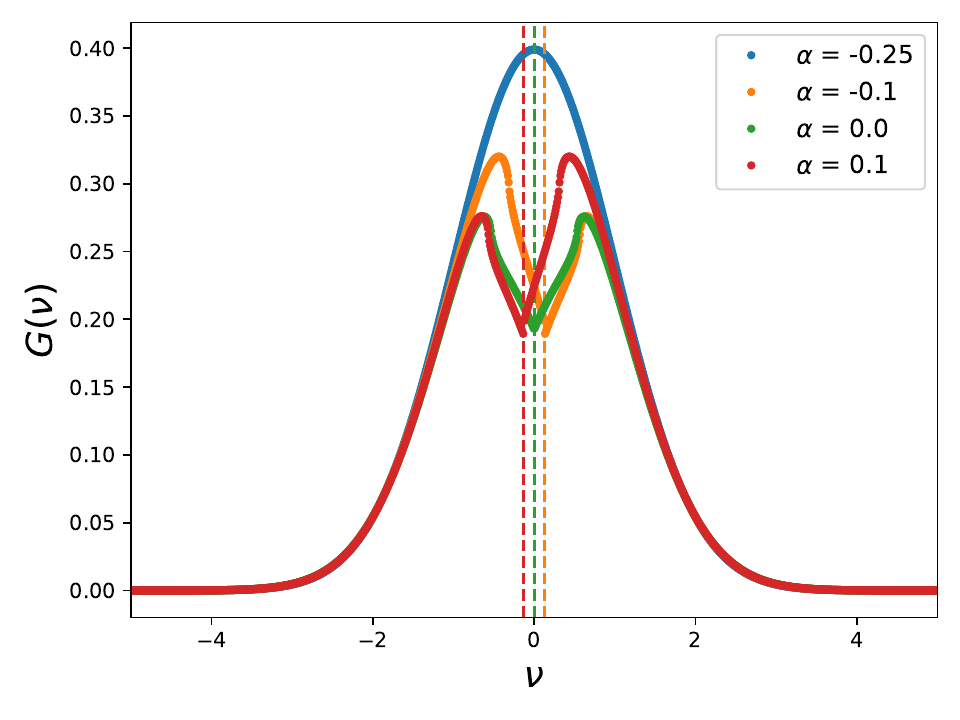}

}\subfloat[$K=1.64$]{\includegraphics[width=0.45\linewidth]{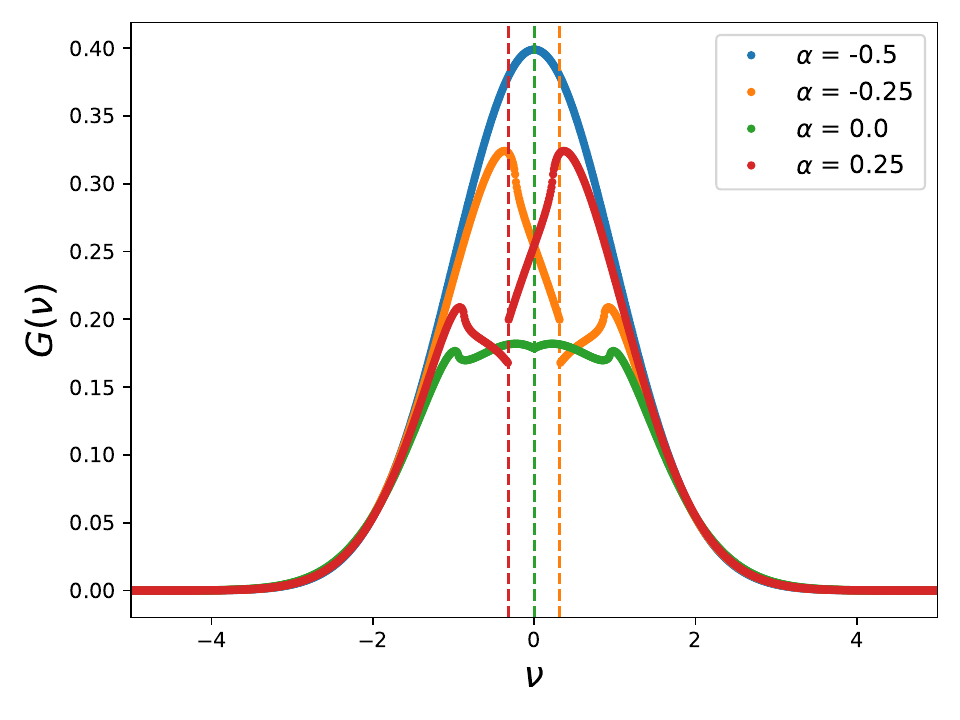}

}
\par\end{centering}
\begin{centering}
\subfloat[$K=1.67$]{\includegraphics[width=0.45\linewidth]{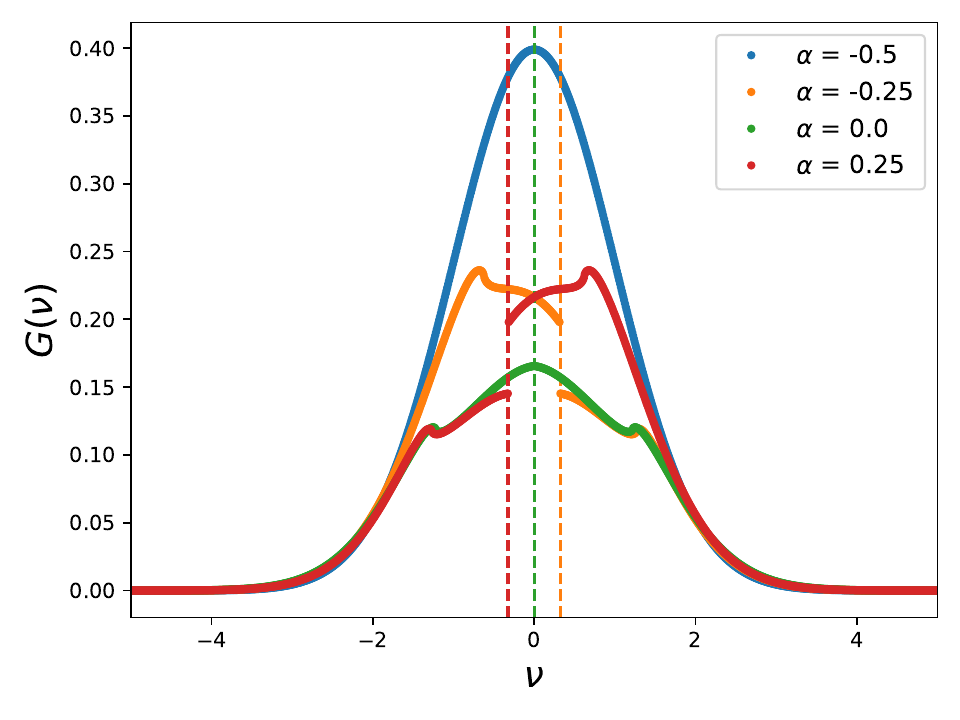}

}\subfloat[$K=1.80$]{\includegraphics[width=0.45\linewidth]{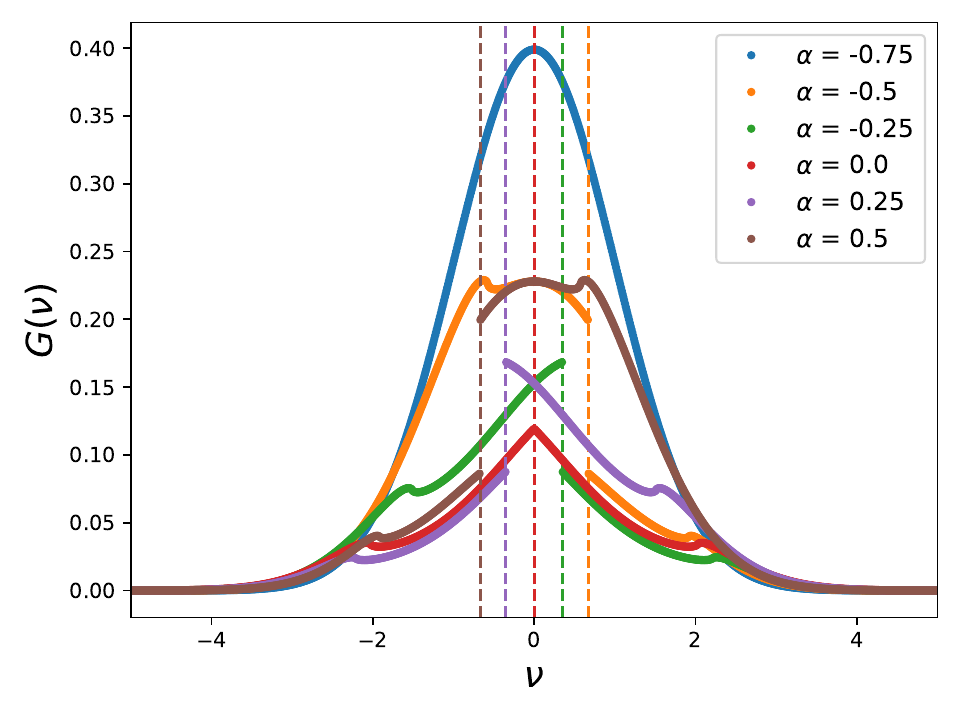}

}
\par\end{centering}
\caption{(a) Fraction of synchronized oscillators, $S(K,\alpha)$. (b-e) Graphs
of $g$ (in blue) and $G_{D}$. As in Fig. \ref{fig:G-Kur-Gauss}(a),
the dashed vertical lines represent the delta term of $G$.}
\label{fig:Gd-K} 
\end{figure}

In this section, we illustrate our analytical result assuming that
natural frequencies are distributed according to the standard normal
density (See Eq. (\ref{eq:normal-density})). Since a complete plot
of $G$ cannot be drawn due to the delta term, we show in Figs. \ref{fig:G-Kur-Gauss}
and \ref{fig:Gd-K} only graphs of $S(K,\alpha)$ and $G_{D}$. The
graphs come from Eqs. (\ref{eq:S-1}) and (\ref{eq:Gd-2}). Each graph
consists of $250$ points.

Fig. \ref{fig:G-Kur-Gauss}(a) shows graphs of $G_{D}$ with $\alpha=0$.
This is the case for which the KS model reduces to the Kuramoto model.
Each graph is plotted with $K$ fixed. The dashed vertical line points
out the delta-term position, given by $\Omega=0$. The area below
the curve connecting neighboring points of a graph of $G_{D}$ corresponds
to the fraction of D oscillators. Fig. \ref{fig:G-Kur-Gauss}(a) indicates
that the fraction of D oscillators diminishes as $K$ increases.

In Fig. \ref{fig:G-Kur-Gauss}(b), we show how $S(K,0)$, the fraction
of S oscillators for $\alpha=0$, varies with $K$. For values of
$K$ less than a critical value $K_{c}\simeq1.5957$\footnote{We calculated the critical value, $K_{c}$, using the formula $K_{c}=\frac{2}{\pi g(\Omega)}$
(See Ref. \cite{daFonseca18} for details). An approximate value for
$K_{c}$ can also be determined from Eq. (\ref{eq:RO-1}) with $\alpha=0$. }, $S(K,0)=0$, i.e. oscillators are all of D type. This is consistent
with Fig. \ref{fig:ROGrid}(a) and Eq. (\ref{eq:S-1}): for $K<K_{c}$
and $\alpha=0$, we have $R=0$, and then the integration range of
Eq. (\ref{eq:S-1}) collapses, leading to $S(K,0)=0$. Note also that,
in Fig. \ref{fig:G-Kur-Gauss}(a), for $K=1.596$, which is slightly
greater than $K_{c}$, the graph of $G_{D}$ resembles, except near
$\nu=0$, the profile of the standard normal density. As $K$ increases
beyond $K_{c}$, Fig. \ref{fig:G-Kur-Gauss}(b) shows an increase
of $S(K,0)$.

The plots in Figs. \ref{fig:Gd-K}(a) show the effects of the parameter
$\alpha$ on $S(K,\alpha)$. Each graph is plotted with $K$ fixed
and $\alpha$ varying in the interval $[-1.0,+1.0]$. For higher values
of $K$, $S(K,\alpha)$ has non-zero values over wider ranges of $\alpha$,
and the maximum points of $S(K,\alpha)$ are positioned at higher
levels. The graphs of $S(K,\alpha)$ are similar to those of $R$,
shown in Fig. \ref{fig:ROGrid}(c): if $R=0$, then $S(K,\alpha)=0$,
as expected from Eq.(\ref{eq:S-1}); for $R>0$, the variation of
$S(K,\alpha)$ follows closely that of $R$, suggesting a monotonic
dependence of $S(K,\alpha)$ on $R.$

Graphs of $G_{D}$ and $g$ are shown in Figs. \ref{fig:Gd-K}(b-e)
for different values of $K$ and $\alpha$. Given $K$ and $\alpha$,
if $R>0$, we show the corresponding $G_{D}$ graph and dashed line
with the same color. If $R=0$ for a given pair $(K,\alpha)$, a graph
of $g$, instead of a $G_{D}$ graph, is shown (with blue dots) to
illustrate the profile of the instantaneous-frequency distribution
in the incoherent state. The set of $K$ values used is the same as
the one of Fig. \ref{fig:Gd-K}(a).

Figures \ref{fig:Gd-K}(b-e) show that, for $\alpha\neq0$, $G_{D}$
has non-symmetric profiles. Yet, for $\alpha=0$, $G_{D}$ profiles
are symmetric (See Fig. \ref{fig:G-Kur-Gauss}(a)). As discussed in
Ref. \cite{Fonseca20}, if $g$ has a profile with a symmetry axis
(e.g. the standard normal density), $G_{D}$ profiles have the same
symmetry axis in the Kuramoto model (KS model with $\alpha=0$). Another
property of the graphs in Figs. \ref{fig:Gd-K}(b-e) is that, for
$\alpha\neq0$, the dashed-line position also differs from zero. The
dashed-line position, given by $\Omega$, has a sign opposite to the
sign of $\alpha$, as shown in Fig. \ref{fig:ROGrid}(d).

A remarkable feature of the graphs in Figs. \ref{fig:Gd-K}(b-e) is
the discontinuity at $\nu=\text{\ensuremath{\Omega}}$. For finite
$\alpha$, the left and right non-symmetric branches of the graphs
are visibly disconnected. For $\alpha=0$, the symmetric branches
\emph{seem} connected at $\nu=\Omega=0$, but, from Eq. (\ref{eq:Gd-2}),
we know that $G_{D}(0)=0$, and the graphs indicate that $G_{D}(0^{+})=G_{D}(0^{-})>0$.
Thus, a discontinuity \emph{seems} to occur even in the symmetric
profiles.

\section{Numerical analysis \label{sec:num}}

\begin{figure}
\begin{centering}
\subfloat[$\alpha=-0.50$]{\includegraphics[width=0.45\linewidth]{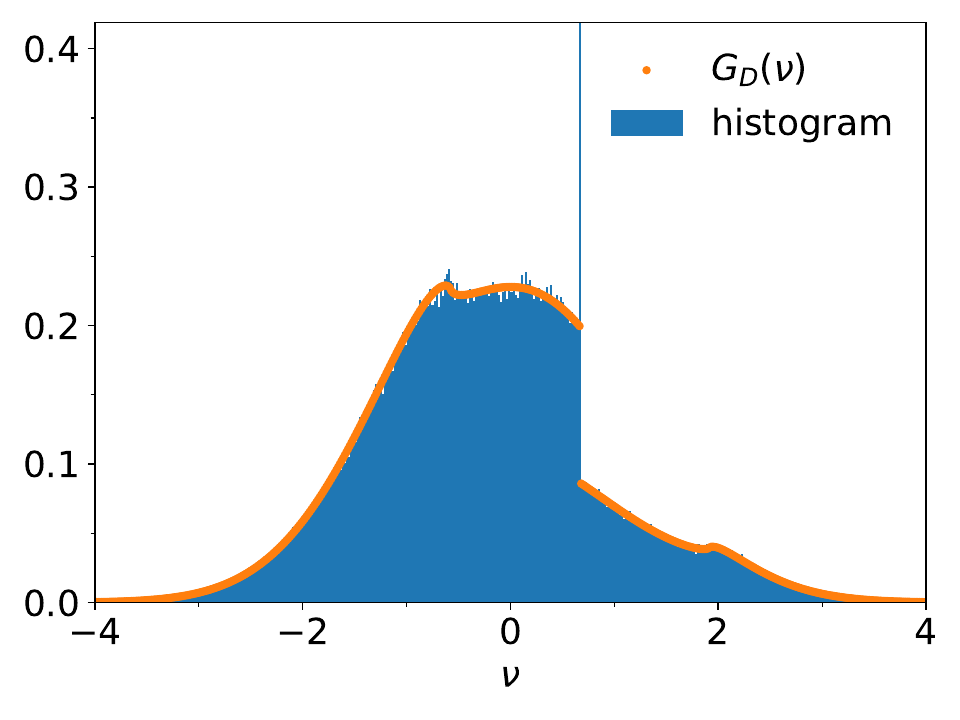}

}\subfloat[$\alpha=-0.25$]{\includegraphics[width=0.45\linewidth]{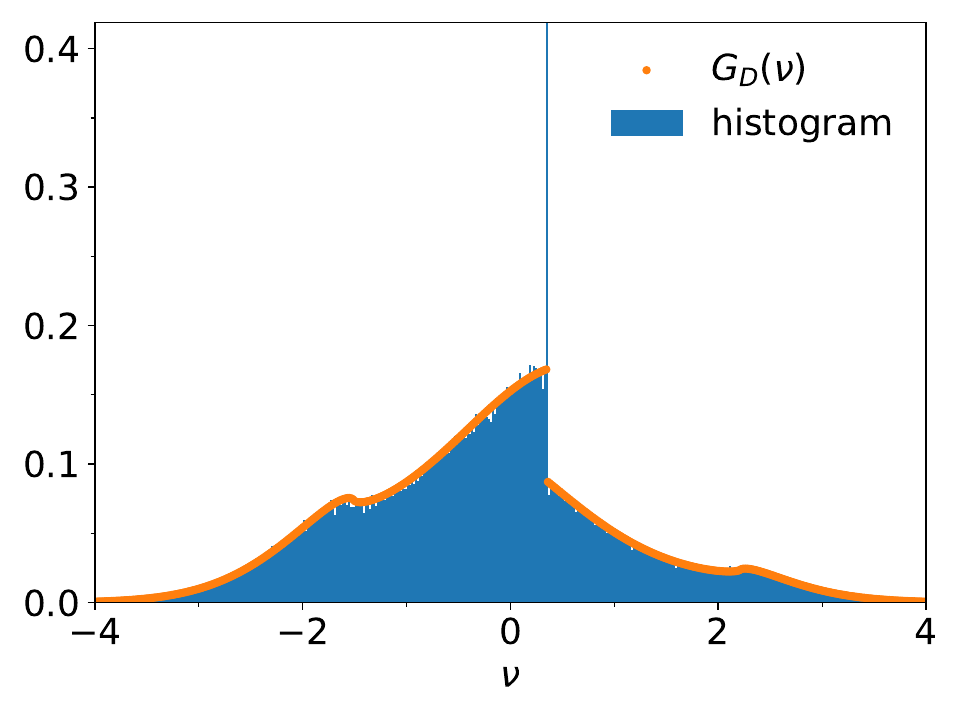}

}
\par\end{centering}
\begin{centering}
\subfloat[$\alpha=+0.25$]{\includegraphics[width=0.45\linewidth]{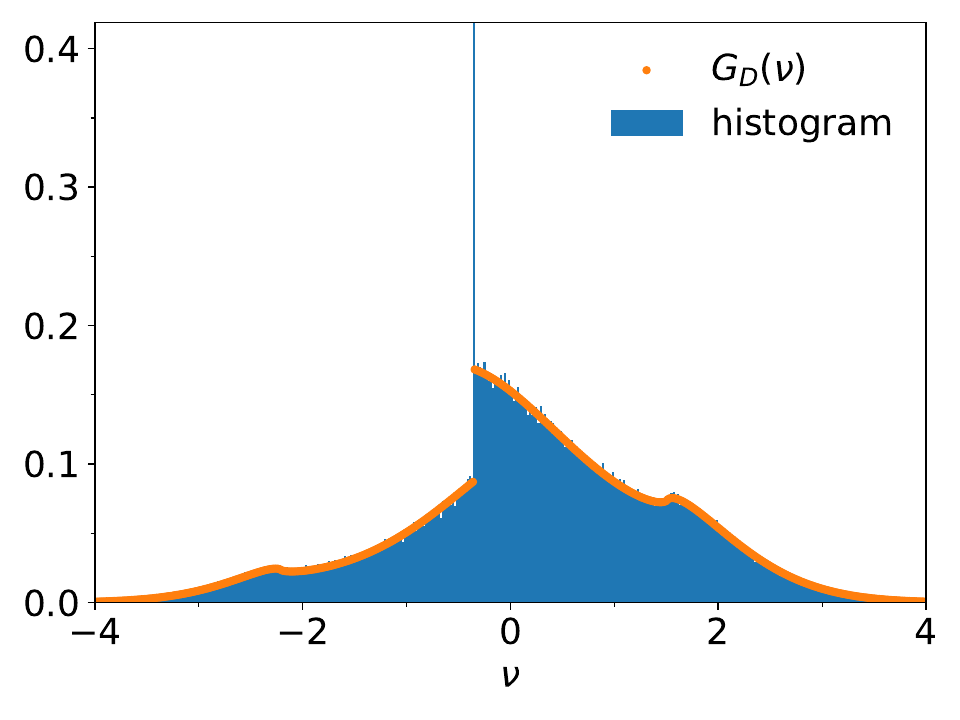}

}\subfloat[$\alpha=+0.50$]{\includegraphics[width=0.45\linewidth]{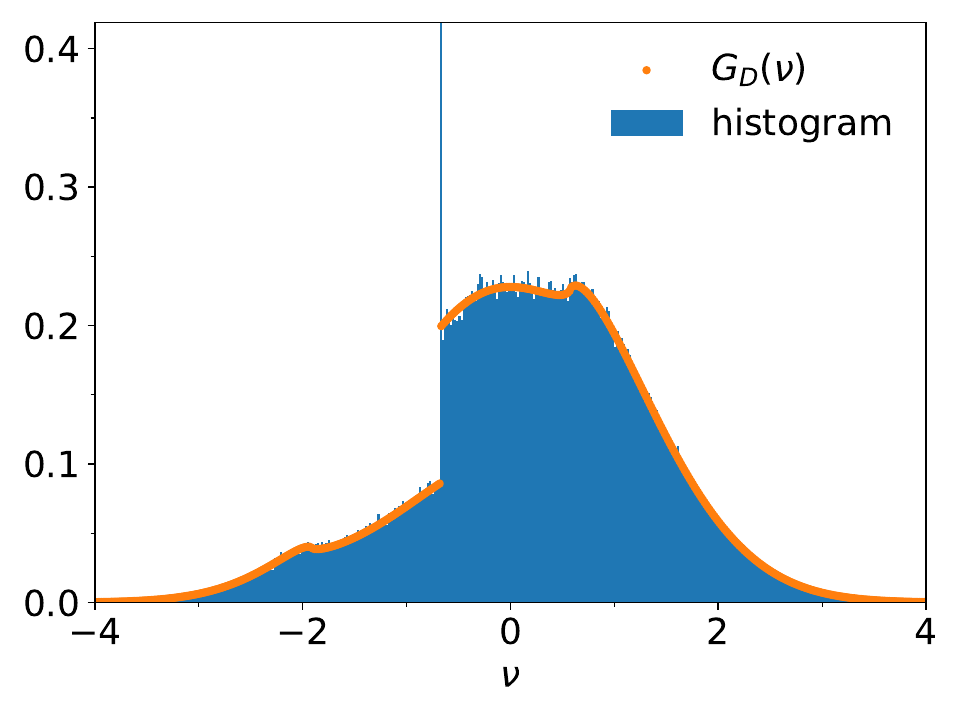}

}
\par\end{centering}
\caption{In blue: normalized histograms of instantaneous frequencies obtained
in numerical simulations of the Kuramoto-Sakaguchi model. In orange:
graphs of $G_{D}$. For all simulations and graphs, $K=1.80$. Model
size: $N=5\times10^{5}$. Simulation time: $T=5\times10^{2}$.}
\label{fig:Gnum-Gd-1} 
\end{figure}

\begin{figure}
\begin{centering}
\subfloat[$K=1.67$; $\alpha=-0.50$]{\includegraphics[width=0.45\linewidth]{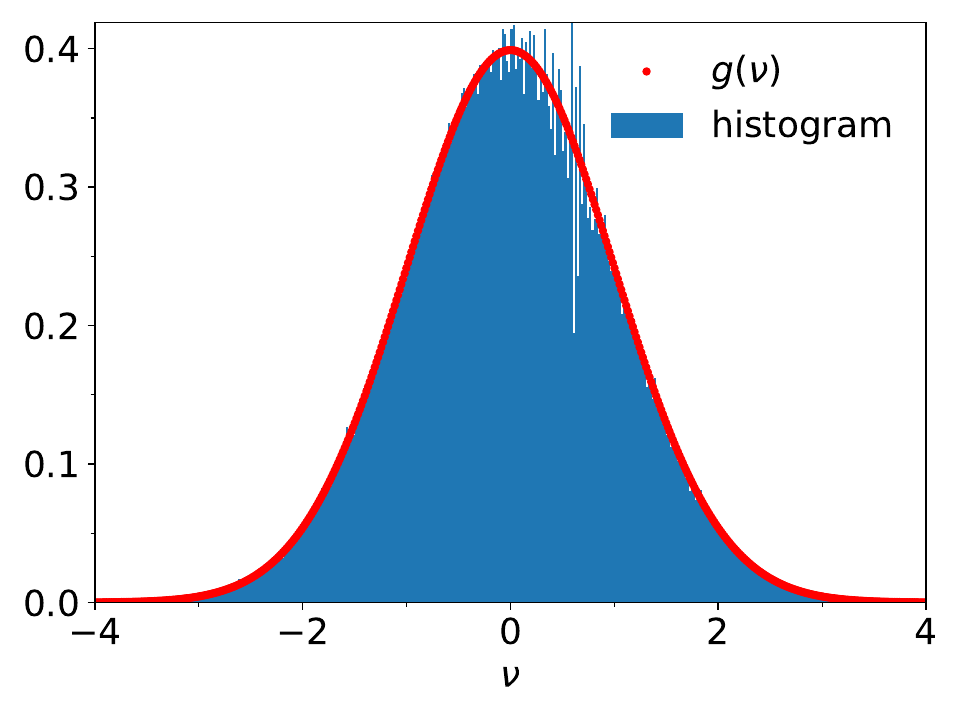}

}\subfloat[$K=1.67$; $\alpha=-0.25$]{\includegraphics[width=0.45\linewidth]{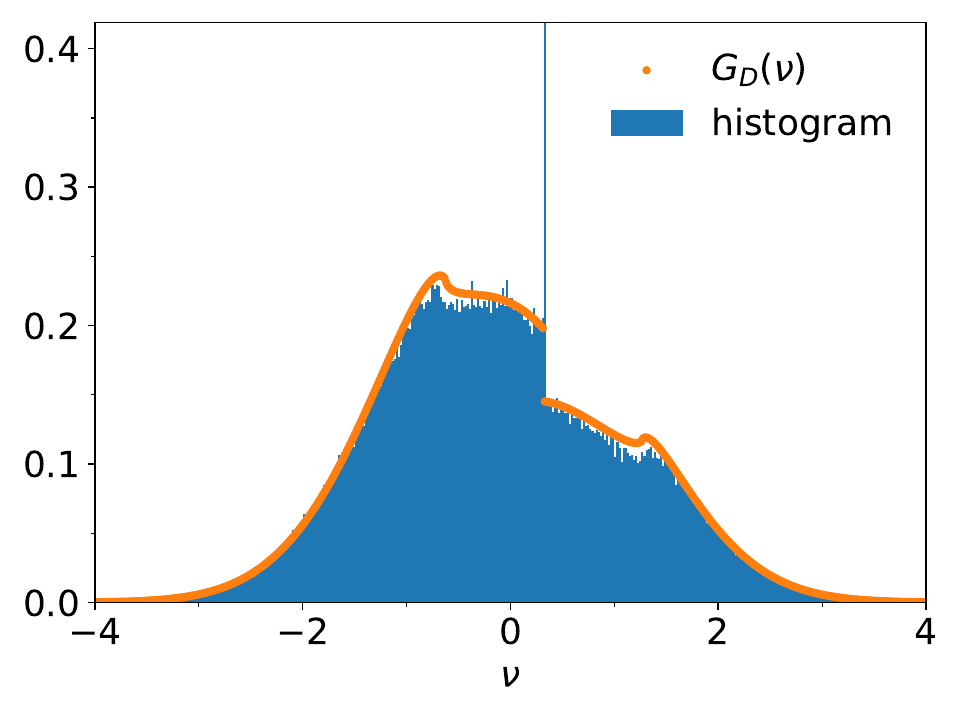}

}
\par\end{centering}
\begin{centering}
\subfloat[$K=1.67$; $\alpha=+0.25$]{\includegraphics[width=0.45\linewidth]{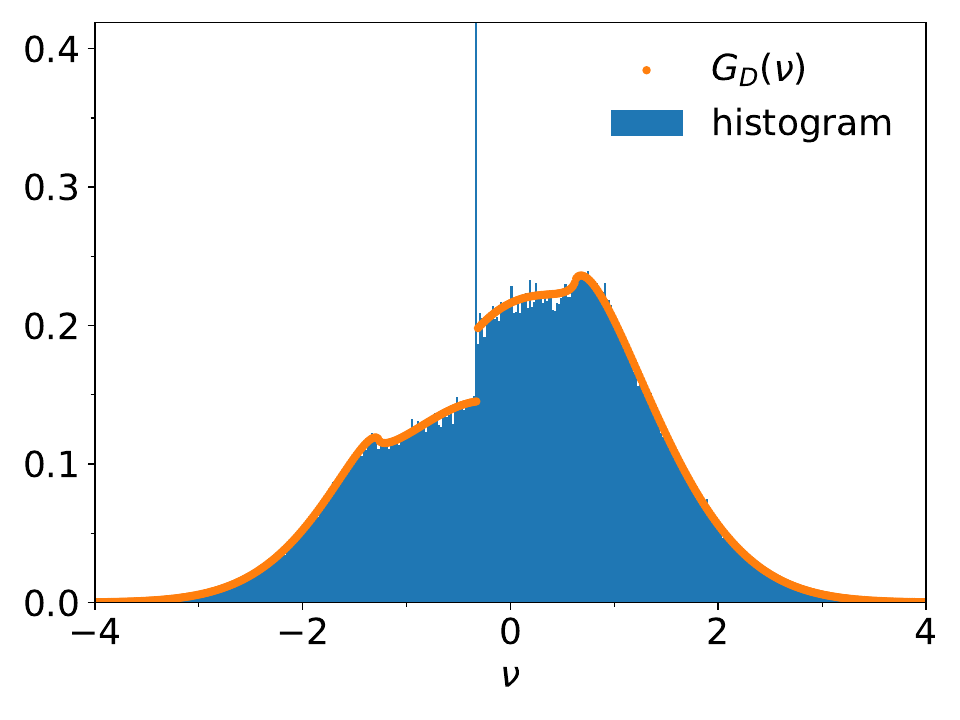}

}\subfloat[$K=1.67$; $\alpha=+0.50$]{\includegraphics[width=0.45\linewidth]{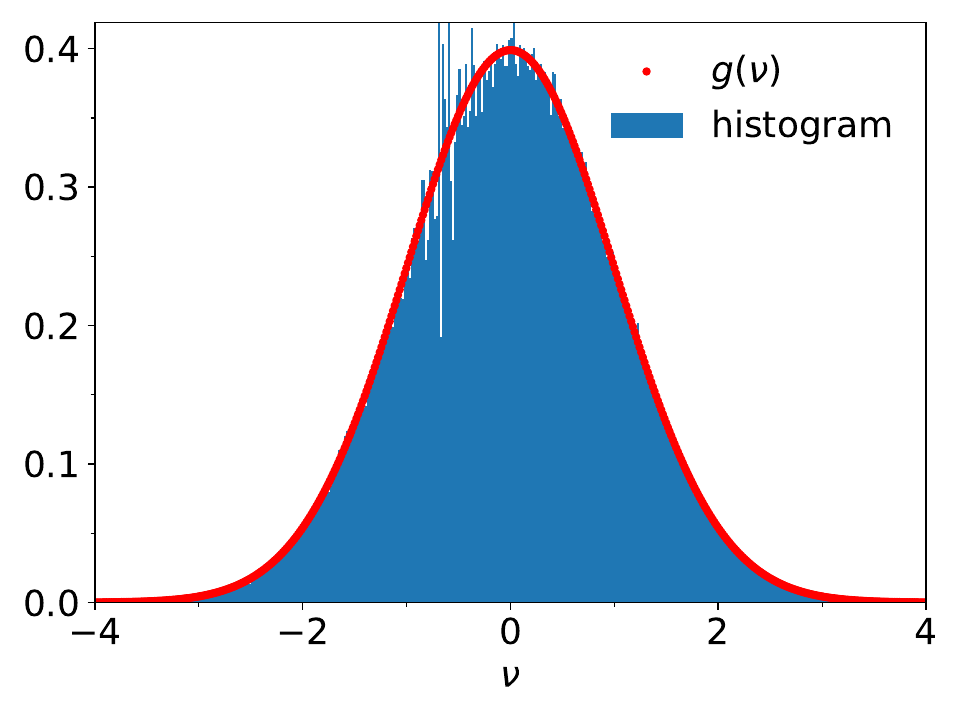}

}
\par\end{centering}
\caption{In blue: normalized histograms of instantaneous frequencies obtained
numerically. In red: graphs of $g$ (a,d). In orange: graphs of $G_{D}$
(b,c). The values of $\alpha$ are the same as in Fig. \ref{fig:Gnum-Gd-1}.
In all simulations and graphs, $K=1.67$. Model size: $N=5\times10^{5}$.
Simulation time: $T=5\times10^{2}$.}
\label{fig:Gnum-Gd-2} 
\end{figure}

In Figs. \ref{fig:Gnum-Gd-1} and \ref{fig:Gnum-Gd-2}, we show graphs
of $G_{D}$ (in orange) and normalized histograms (in blue) of instantaneous
frequencies obtained in numerical simulations of the KS model. By
numerical simulation we mean numerical integration of the $N$ differential
equations of the KS model from the initial time instant $0$ to a
final one $T>0$. All simulations were performed with the numerical
library ODEPACK \cite{Hindmarsh}. The ODEPACK's solver used is \emph{LSODA},
a hybrid implementation of Adams and BDF methods \cite{Petzold}.
Again, we assume a standard normal density of natural frequencies. 

Before initiating a simulation, two random samples are generated:
a sample $\ensuremath{\ensuremath{\left\{ \omega_{i}\right\} _{i=1}^{N}}}$
of random natural frequencies and another one of random initial phases,
$\ensuremath{\left\{ \theta_{i}(0)\right\} _{i=1}^{N}}$. A standard-normal
generator of random number s is used to create the sample of natural
frequencies. The random initial phases are generated according to
a uniform distribution in the interval $-\pi<\theta_{i}(0)<\pi$.

After a simulation is concluded, a normalized histogram (a histogram
with the unit area) with bars of equal width is built from the instantaneous
frequencies $\left\{ \dot{\theta_{i}}(T)\right\} _{i=1}^{N}$, computed
from Eqs. (\ref{eq:KS-model}) and the numerically-obtained set of
phases $\left\{ \theta_{i}(T)\right\} _{i=1}^{N}$. For the histograms
of Figs. \ref{fig:Gnum-Gd-1} and \ref{fig:Gnum-Gd-2}, the simulation
parameters are $N=5\times10^{5}$ and $T=5\times10^{2}$.

The occurrence of both S and D oscillators is depicted in the histograms
of Figs. \ref{fig:Gnum-Gd-1}(a-d), where $K=1.80$, and $\alpha$
takes values in the set $\left\{ -0.5,-0.25,+0.25,+0.5\right\} $.
Figs. \ref{fig:Gnum-Gd-1}(a-d) show that the graphs of $G_{D}$ are
in agreement with the histogram profiles. The histograms exhibit,
at $\nu=\Omega$, tall and thin bars, which we refer to as\emph{ peaks}.
The peaks are related to the delta term of the function $G$ (See
Eq. (\ref{eq:G-1})).

We avoided showing the peaks entirely because their height are much
higher than the other histogram bars. The smaller the width of the
peak, the higher its height. The peaks emerge due to the large accumulation
of instantaneous frequencies, most of them from S oscillators. We
draw the reader's attention to the fact that the area of the peak
is not exactly equal to $S(K,\alpha)$: a small part of the peak area
contributes to the fraction of D oscillators, given by $1-S(K,\alpha)$.
Yet, the peak area can be a good approximation to $S(K,\alpha)$,
depending on how small is the peak width.

A noteworthy property is the seeming reflection symmetry around the
peaks. The reflection symmetry is related to sign inversions in the
parameter $\alpha$: if the signal of $\alpha$ is inverted, keeping
fixed its absolute value, the right (or left) side of the graphs and
histograms are reflected on the left (or right) side.

The same type of symmetry is observed in Figs. \ref{fig:Gnum-Gd-2}(a-d),
where $K=1.60$, and the set of $\alpha$ values is the same as in
Figs. \ref{fig:Gnum-Gd-1}(a-d). In Figs. \ref{fig:Gnum-Gd-2}(a)
and \ref{fig:Gnum-Gd-2}(d) we plot $g$ (in red) instead of $G_{D},$
since $R=0$ (incoherent state) for the values of $K$ and $\alpha$
considered (as indicated in Fig. \ref{fig:ROGrid}(c)). The curves
of $g$ fit histograms showing no clearly visible peaks. A synchronization
state is represented in Figs. \ref{fig:Gnum-Gd-2}(b) and \ref{fig:Gnum-Gd-2}(c).
The curves of $G_{D}$ (in orange) also fit the histograms, but small
deviations occur due to time fluctuations in the histogram bars.

\begin{figure}
\begin{centering}
\subfloat[$N=5\times10^{5}$; $T_{1}=500$]{\includegraphics[width=0.33333\linewidth]{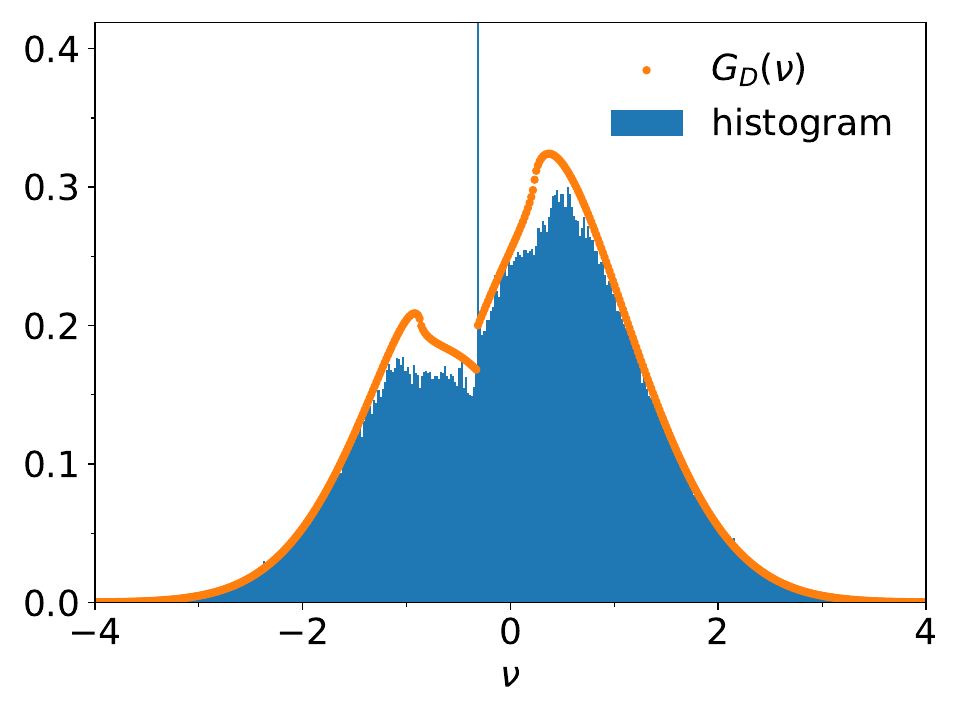}

}\subfloat[$N=5\times10^{5}$; $T_{2}=1000$]{\includegraphics[width=0.33333\linewidth]{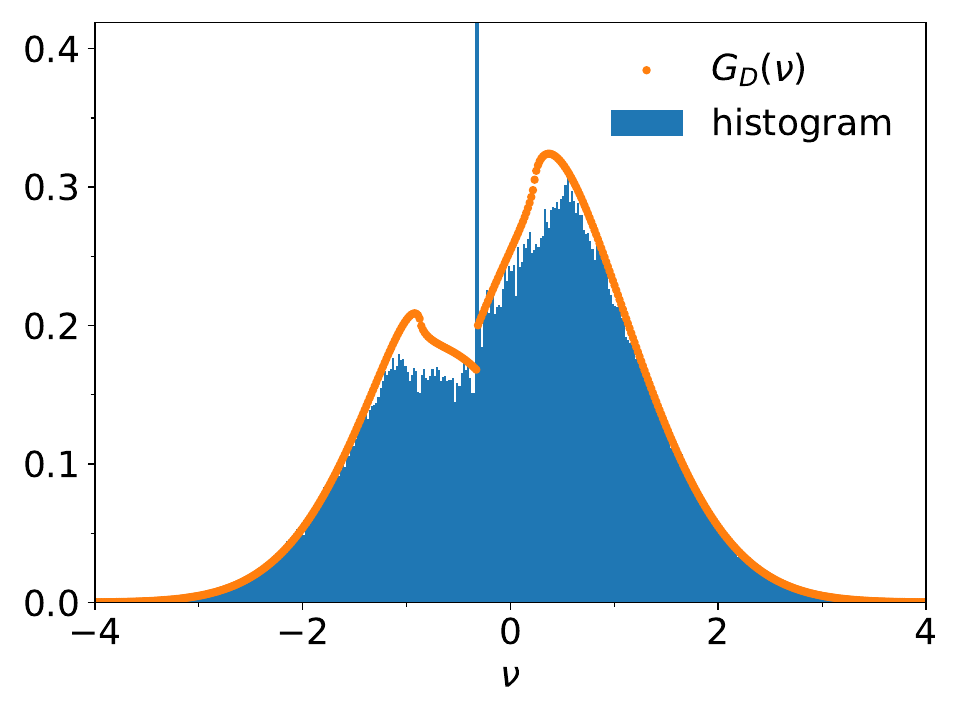}

}\subfloat[$N=5\times10^{5}$; $T_{3}=1500$]{\includegraphics[width=0.33333\linewidth]{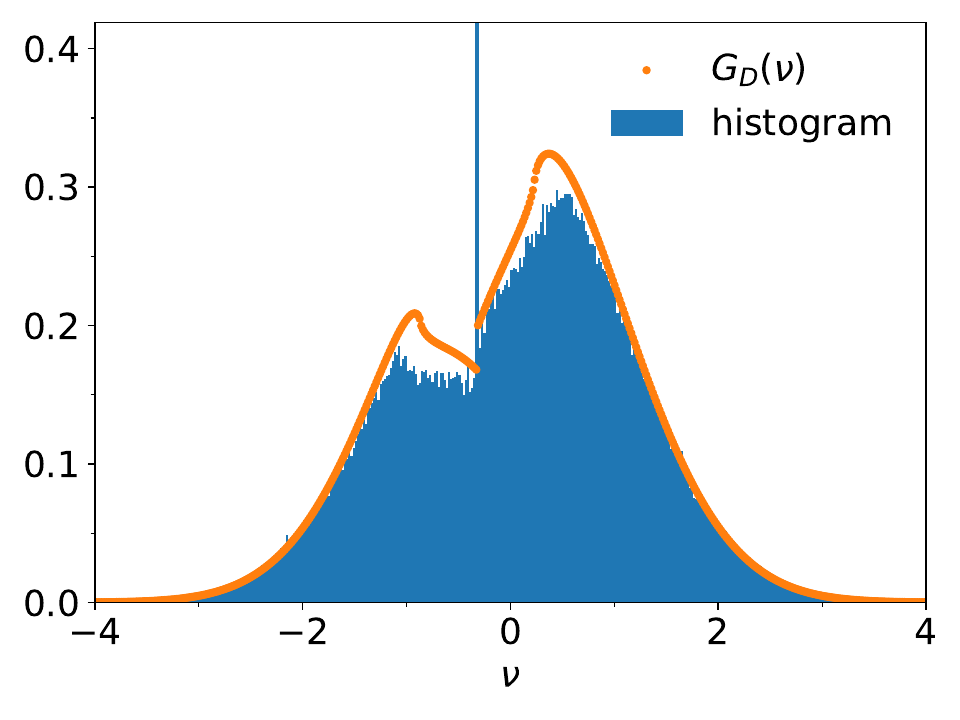}

}
\par\end{centering}
\begin{centering}
\subfloat[$N=1.5\times10^{6}$; $T_{1}=500$]{\includegraphics[width=0.33333\linewidth]{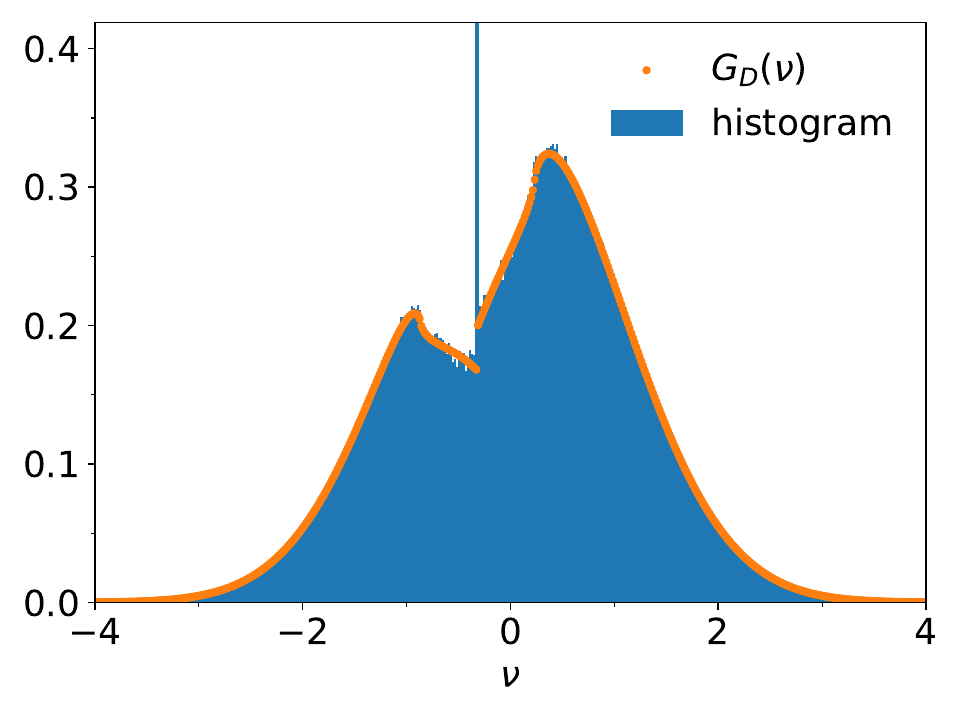}

}\subfloat[$N=1.5\times10^{6}$; $T_{2}=1000$]{\includegraphics[width=0.33333\linewidth]{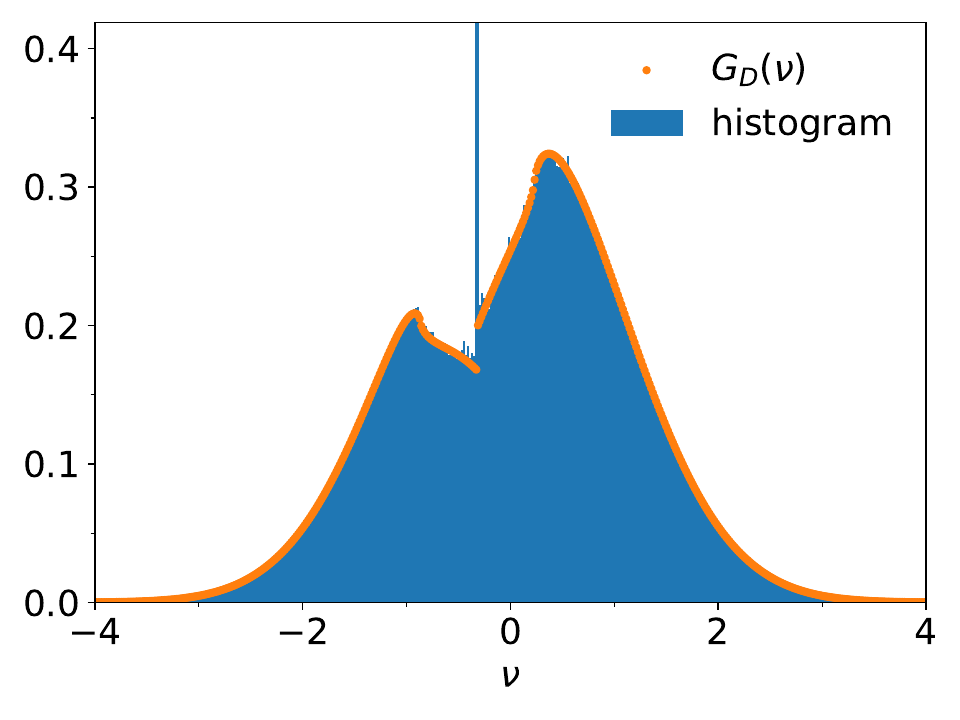}

}\subfloat[$N=1.5\times10^{6}$; $T_{3}=1500$]{\includegraphics[width=0.33333\linewidth]{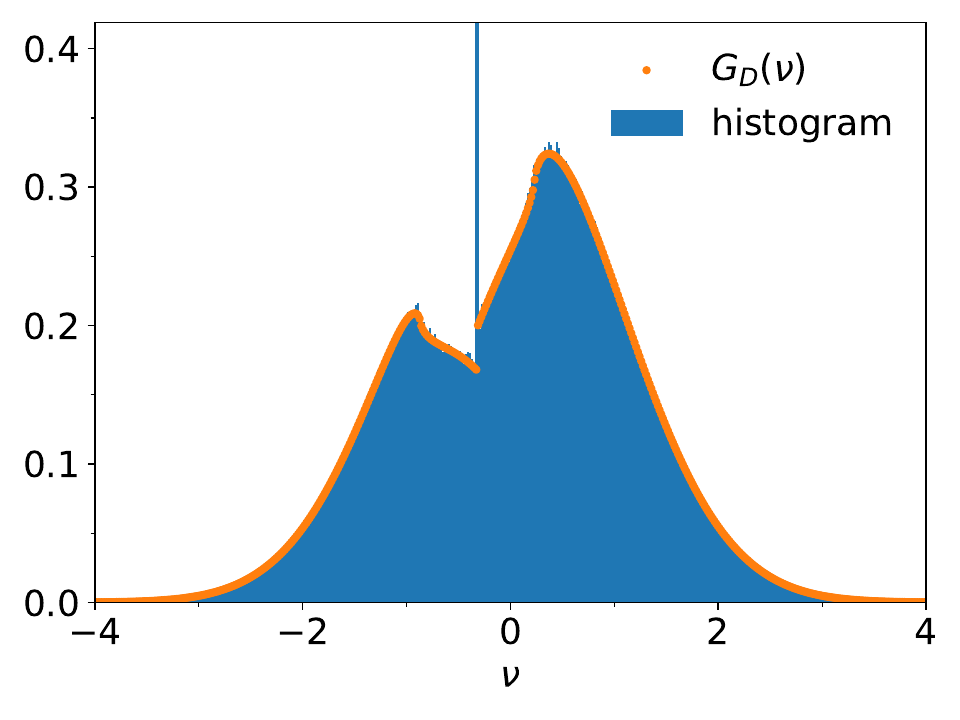}

}
\par\end{centering}
\caption{In blue: normalized histograms of instantaneous frequencies at the
time instants $T_{1}$, $T_{2},$ and $T_{3}$ obtained in two simulations
of the KS model performed in the time interval $\left[0,T_{3}\right]$
and with different numbers of oscillators: one with $N=5\times10^{5}$
and the other with $N=1.5\times10^{6}$. The histograms in (a), (b)
and (c) result from the simulation with the smallest $N$, while those
in (d), (e), and (f), from the one with the highest $N$. For all
simulations and graphs of $G_{D}$ (in orange), $K=1.64$ and $\alpha=0.25$.}
\label{fig:Gnum-Gd-K164} 
\end{figure}

Such fluctuations and deviations are also present in the histograms
of Figs. \ref{fig:Gnum-Gd-K164}(a-c). The deviations occur near the
synchronization peak. The sequence of figures \ref{fig:Gnum-Gd-K164}(a),
\ref{fig:Gnum-Gd-K164}(b) and \ref{fig:Gnum-Gd-K164}(c) depicts
the evolution of the instantaneous-frequency distribution in histograms
from three different time instants of the same simulation: $T_{1}=500$,
$T_{2}=1000$, and $T_{3}=1500$. The simulation time interval is
$\left[0,T_{3}\right]$, and we define $K=1.64$, $\alpha=0.25$,
and $N=5\times10^{5}$. Again, initial phases and natural frequencies
are random numbers generated according to the uniform and standard
normal distributions.

The histograms of Figs .\ref{fig:Gnum-Gd-K164}(d), \ref{fig:Gnum-Gd-K164}(e),
and \ref{fig:Gnum-Gd-K164}(f) comes from a simulation similar to
the previous one. The difference is that the number of oscillators
is three times higher, requiring a new random sample of initial phases
and natural frequencies. The histograms show a more stationary profile
in the sequence of time instants $T_{1}$, $T_{2},$ and $T_{3}$.
In addition, the graphs of $G_{D}$ are in better agreement with the
histogram profiles. 

\section{Conclusions\label{sec:Conclusion}}

In this work, we showed how to obtain the density of instantaneous
frequencies in the Kuramoto-Sakaguchi model. The density of instantaneous
frequencies is a stationary probability density function with a complex
formula given by the sum of two terms: a Dirac-delta-type function
and a discontinuous one. 

The Dirac-delta term is located in the synchronization frequency and
carries information about the number of synchronized oscillators.
The other term is discontinuous at a point defined also by the synchronization-frequency
value. The discontinuous term have profiles with varied and unexpected
shapes, and the area below them gives the fraction of out-of-synchrony
oscillators.

Our formula is a generalization of the one obtained in Ref. \cite{Fonseca20}
for the \emph{Kuramoto} model (i.e. Kuramoto-Sakaguchi model with
a phase-lag equal to zero). The formulas are mathematically quite
similar, particularly concerning the property that the generalization
has no explicit dependence on the phase-lag parameter. Indeed, the
dependence on the phase-lag is implicit and takes place through the
order parameter and the synchronization frequency.

Contrary to what was shown for the Kuramoto model in Ref. \cite{Fonseca20},
natural-frequency densities with a symmetry axis in the synchronization
frequency does\emph{ not }imply the same type of symmetry in the density
of instantaneous frequencies for the Kuramoto-Sakaguchi model. However,
this density exhibits a reflection symmetry, characterized by a flip
of the density profile induced by sign inversions in the phase-lag
parameter.

Our result is in accordance with numerical simulations of the Kuramoto-Sakaguchi
model, provided simulations are performed with a large enough number
of oscillators. If the number of oscillators is such that the instantaneous-frequency
distribution from simulations (normalized histograms) shows significant
non-stationary behavior, one observes a less robust fit to simulation
data. But increasing the number of oscillators suppresses the non-stationarity
and improves the quality of the theoretical prediction. These finite-size
effects, also analyzed in the context of the Kuramoto model \cite{Fonseca20},
are expected to occur in the Kuramoto-Sakaguchi model. Our result
is based on the Kuramoto-Sakaguchi theory, which includes equilibrium
assumptions devised under the requirement of an infinite number of
oscillators and infinitely long times.

New research directions can be taken with this work as a starting
point. A mathematically-oriented subject would be finding asymptotic
analytical properties of the density function near the synchronization
frequency as well as in its tails. Extending the study presented here,
considering other types of natural-frequency densities, including
non-symmetric and non-unimodal ones, is also an interesting topic.
In addition, our formula can be used to determine the instantaneous-frequency
statistical moments, which can then be compared to the moments of
the natural frequencies and used to compute relevant quantities such
as expected values and variances.

More importantly, we envision a non-stationary model capturing the
collective dynamics of the instantaneous frequencies. We conjecture
that the model solution would be a time-dependent density of instantaneous
frequencies, and our formula would reflect the solution behavior in
the long-time limit.

\section*{Acknowledgments}

This work was made possible through financial support from Brazilian
research agency FAPESP (grant n. 2019/12930-9). JDF thanks Hugues
Chaté for valuable discussions. EDL thanks support from Brazilian
agencies CNPq (301318/2019-0) and FAPESP (2019/14038-6).

\end{document}